\documentclass[amsmath,amssymb,aps,pra,superscriptaddress,twocolumn
]{revtex4-2}

\usepackage{amssymb}
\usepackage{graphicx}
\usepackage{subcaption} 
\usepackage{dcolumn}
\usepackage{bm}
\usepackage{amsmath}
\usepackage[colorlinks,urlcolor=cyan,citecolor=blue,linkcolor=magenta]{hyperref}
\usepackage{ragged2e}

\begin{document}
\title{Breathing mode of quantum droplets in dipolar quantum gases: A sum-rule analysis}

\affiliation{Zhejiang Key Laboratory of Quantum State Control and Optical Field Manipulation, Department of Physics, Zhejiang Sci-Tech University, Hangzhou 310018, China}
\affiliation{Lanzhou Center for Theoretical Physics, Key Laboratory of Theoretical Physics of Gansu Province, Key Laboratory of Quantum Theory and Applications of MoE, Gansu Provincial Research Center for Basic Disciplines of Quantum Physics, Lanzhou University, Lanzhou 730000, China}

\author{Xinran Zhang}
\affiliation{Zhejiang Key Laboratory of Quantum State Control and Optical Field Manipulation, Department of Physics, Zhejiang Sci-Tech University, Hangzhou 310018, China}

\author{Junli Liu}
\affiliation{Zhejiang Key Laboratory of Quantum State Control and Optical Field Manipulation, Department of Physics, Zhejiang Sci-Tech University, Hangzhou 310018, China}

\author{Huiyun Xiao}
\affiliation{Zhejiang Key Laboratory of Quantum State Control and Optical Field Manipulation, Department of Physics, Zhejiang Sci-Tech University, Hangzhou 310018, China}

\author{Xiao-Long Chen}\email{xiaolongchen@zstu.edu.cn}
\affiliation{Zhejiang Key Laboratory of Quantum State Control and Optical Field Manipulation, Department of Physics, Zhejiang Sci-Tech University, Hangzhou 310018, China}
\affiliation{Lanzhou Center for Theoretical Physics, Key Laboratory of Theoretical Physics of Gansu Province, Key Laboratory of Quantum Theory and Applications of MoE, Gansu Provincial Research Center for Basic Disciplines of Quantum Physics, Lanzhou University, Lanzhou 730000, China}

\author{Yunbo Zhang}
\email{ybzhang@zstu.edu.cn}
\affiliation{Zhejiang Key Laboratory of Quantum State Control and Optical Field Manipulation, Department of Physics, Zhejiang Sci-Tech University, Hangzhou 310018, China}

\date{\today}

\begin{abstract}
We investigate theoretically the ground-state properties and breathing-mode collective excitations of three-dimensional dipolar Bose gases in anisotropic harmonic traps incorporating quantum fluctuations. Combining a Gaussian variational ansatz with a sum-rule analysis, we derive explicit analytical expressions for both axial and radial breathing-mode frequencies, which are validated by numerical solutions of the time-dependent extended Gross-Pitaevskii equation. Our theoretical predictions show great agreement with existing experimental data for $^{166}$Er and $^{162}$Dy gases. By constructing comprehensive phase diagrams across the parameter space of the $s$-wave scattering length, atom number, and trap aspect ratio, we reveal both discontinuous first-order phase transitions and smooth crossovers between the dilute Bose-Einstein condensate and dense quantum droplet phases. We confirm that the enhanced incompressibility induced by quantum fluctuations significantly elevates the breathing-mode frequencies in the droplet phase compared with conventional weakly interacting Bose gases.
Furthermore, the system undergoes a phase transition and a crossover over the scattering length under the elongated and flattened confinements, characterized by discontinuous jumps and continuous crossovers in peak density and atomic cloud sizes, respectively.
Our work offers a useful analytical framework to characterize collective excitations in dipolar quantum gases, providing quantitative insights for forthcoming ultracold atom experiments in lanthanide atoms and polar molecules.
\end{abstract}

\maketitle

\section{INTRODUCTION}

Dipolar Bose-Einstein condensates (BECs) of magnetic atoms such as chromium ($^{52}\mathrm{Cr}$)~\cite{griesmaier2005bose}, dysprosium ($^{162}\mathrm{Dy}$, $^{164}\mathrm{Dy}$)~\cite{lu2011strongly}, and erbium ($^{166}\mathrm{Er}$, $^{168}\mathrm{Er}$)~\cite{aikawa2012bose} provide an ideal platform for exploring quantum many-body physics with long-range and anisotropic dipole-dipole interactions (DDIs)~\cite{lahaye2009physics,baranov2012condensed,bottcher2021new,luo2021new,guo2021new,norcia2021developments,chomaz2023dipolar}. Unlike conventional atomic gases dominated by isotropic short-range contact interactions, dipolar condensates exhibit a rich interplay between mean-field repulsion and DDI attraction, leading to a variety of exotic phenomena. In particular, when the repulsive contact interaction and the attractive dipolar interaction nearly cancel each other, the standard mean-field theory predicts mechanical instability. However, the system instead does not collapse and stabilizes into a novel self-bound state known as a quantum droplet.
This stabilization is fundamentally due to the beyond-mean-field repulsive Lee-Huang-Yang (LHY) correction induced by quantum fluctuations~\cite{schutzhold2006mean,lima2011quantum,lima2012beyond}, which adds an energy density scaling as $n^{5/2}$, i.e., more sensitive to density than the mean-field energy (i.e., $\propto n^2$).
The competition and eventual balance between the repulsive mean-field and LHY contributions over the attractive DDI suppresses the collapse, allowing the formation of ultradilute yet robust droplets even in free space. This stabilization mechanism of high-order LHY energy induced by quantum fluctuations was first predicted by Petrov in Bose-Bose mixtures~\cite{petrov2015quantum}, and these self-bound droplet states were soon observed in dipolar quantum gases such as $^{164}$Dy~\cite{kadau2016observing,ferrier2016observation,schmitt2016self} and $^{166}$Er~\cite{chomaz2016quantum}, followed by studies in $^{162}$Dy~\cite{tanzi2019supersolid,bottcher2019dilute,tanzi2019observation}. 
Since then, quantum droplets have also been observed in homonuclear Bose–Bose mixtures (e.g., $^{39}$K~\cite{cabrera2018quantum,semeghini2018self,cheiney2018bright,ferioli2019collisions,skov2021observation}) and heteronuclear mixtures (e.g., $^{41}$K-$^{87}$Rb~\cite{derrico2019observation,burchianti2020dual,cavicchioli2025dynamical}, $^{23}$Na-$^{87}$Rb~\cite{guo2021lhy}). It ij worth mentioning that, very recently by utilizing the microwave shielding technique, experimental physicists have realized BECs using ultracold sodium-cesium (NaCs) polar molecules with relatively strong DDI between molecules~\cite{bigagli2024observation}, and the self-bound droplets have also been successfully observed in these ultracold dipolar molecules~\cite{zhang2026observation}.

In dipolar atomic gases, the system can exhibit diverse quantum-fluctuation-stabilized droplet and density-ordered phases depending on the anisotropic DDI strength, atom number, and confinement geometry. As the DDI strength increases, the effectively attractive mean-field interaction is counterbalanced by the repulsive LHY correction, which prevents collapse and leads to dense quantum droplets. When the DDI strength continues to increase, roton softening (instability) may induce spontaneous density modulation, i.e., supersolid-like states and ordered droplet arrays~\cite{wenzel2017striped,bottcher2019transient,chomaz2019long,tanzi2019supersolid,norcia2021two,zhen2025breaking}. In particular, the supersolid-like phase simultaneously breaks the continuous gauge U(1) symmetry (superfluidity) and the continuous spatial translational symmetry (crystal pattern), and is therefore closely associated with the long-sought supersolid phase of matter.
Beyond uniform droplets and ordered arrays, investigations have been extended to vortex-carrying quantum droplets~\cite{klaus2022observation,casotti2024observation,li2024strongly}, multicomponent or asymmetric dipolar mixtures~\cite{trautmann2018dipolar, durastante2020feshbach, politi2022interspecies}, etc.
Following the successful synthesis of stable dipolar quantum droplets, experimental research has rapidly shifted toward probing their collective excitations and dynamic properties in magnetic quantum gases~\cite{chomaz2023dipolar}. Initial experimental efforts focused on measuring the lowest monopole (breathing) mode in erbium atoms across the smooth crossover from a dilute BEC to a dense droplet~\cite{chomaz2016quantum}, which is emphasized in this work. Meanwhile, collision experiments between dipolar droplets have also been conducted to probe the dynamical properties of the system~\cite{ferrier2016liquid}, followed by the observation of the scissors mode in dipolar quantum droplets of a single dysprosium gas~\cite{ferrier2018scissors}.

From a theoretical perspective, several effective frameworks have been developed to study this quantum-fluctuation-stabilized dipolar droplet, including the widely used extended Gross-Pitaevskii equation (GPE), quantum Monte Carlo simulations~\cite{saito2016path,cinti2017superfluid,cinti2017classical,bombin2017dipolar,bottcher2019dilute}, the Bogoliubov-de Gennes equations \cite{ronen2007radial,wilson2010critical,baillie2017collective,roccuzzo2019supersolid,hertkorn2019fate,pal2022infinite,blakie2023axial,schubert2026soliton}, and Hartree–Fock–Bogoliubov theories~\cite{boudjemaa2018fluctuations,aybar2019temperature,boudjemaa2020quantum}.
The extended GPE incorporates the LHY correction under the local density approximation and has proven highly successful in describing the static and dynamic properties of quantum droplets~\cite{baillie2016self,bisset2016ground,wachtler2016ground,wachtler2016quantum,schmitt2016self,chomaz2016quantum,bottcher2019dilute,tanzi2019supersolid,tanzi2019observation,bisset2021quantum,smith2021quantum}. However, full three-dimensional time-dependent extended GPE calculations are computationally expensive, motivating the development of simpler analytical approaches. The variational method provides an intuitive and efficient alternative for mapping out phase diagrams and ground-state properties~\cite{santos2000bose,yi2001trapped,wachtler2016ground,bisset2016ground,saito2016path,cidrim2018vortices}.
For low-energy collective excitations, the sum-rule approach based on linear response theory provides estimated upper bounds for collective mode frequencies directly from ground-state expectation values~\cite{pitaevskii2016bose,menotti2002collective,ferrier2018scissors,de2021polarization,blakie2023axial,orignac2024breathing}, without solving the full Bogoliubov–de Gennes equations. In particular, monopole (breathing) modes are exceptionally sensitive to the compressibility of the macroscopic wave function, making them ideal for characterizing the transition from dilute condensates to self-bound droplets~\cite{chomaz2016quantum,tanzi2019supersolid,zhen2025breaking}. These developments highlight the need for accurate theoretical descriptions of collective excitations, which serve as efficient probes of phases of matter and phase boundaries.

In this work, motivated by these open questions, we systematically investigate the low-energy collective excitations of three-dimensional dipolar quantum gases, especially the quantum droplets, confined in anisotropic harmonic traps. While prior variational works have focused predominantly on stationary properties, we construct such an analytical framework by combining the Gaussian variational ansatz with a nonperturbative sum-rule analysis, and derive explicit formulas for both axial and radial breathing-mode frequencies. These analytical predictions are then validated by direct numerical simulations of the time-dependent extended GPE and compared with existing experimental measurements across the droplet-BEC crossover in $^{166}$Er~\cite{chomaz2016quantum} and within the BEC regime in $^{162}$Dy~\cite{tanzi2019supersolid}. Using these expressions, we construct phase diagrams in the parameter space of scattering length, atom number, and trap aspect ratio, revealing first-order phase transitions and continuous crossovers between the low-density BEC and high-density droplet phases. We further investigate the role of trap anisotropy, focusing on elongated and flattened limits, and demonstrate that the breathing-mode frequency is significantly enhanced in the droplet regime due to the increased incompressibility from quantum fluctuations.

The remainder of this paper is organized as follows: In Sec.~\ref{sec:theory}, we present the theoretical framework, detailing the energy functional of a dipolar quantum gas including quantum fluctuations, the Gaussian variational approach, and the sum-rule analysis for breathing-mode frequencies, followed by a brief overview of the extended GPE used for numerical validation. In Sec.~\ref{sec:results}, we present and discuss our main results, beginning with typical phase diagrams and the characterization of discontinuous phase transitions and smooth crossovers. We then quantitatively compare our theoretical predictions of mode frequencies with recent experimental measurements for $^{166}$Er and $^{162}$Dy atoms, examine the impact of anisotropic trap confinement, and evaluate the specific limits of elongated and flattened configurations. Finally, Section~\ref{sec:conclusions} summarizes our conclusions and provides an outlook for future research.

\section{THEORETICAL FRAMEWORK}\label{sec:theory}

In this section, we present two key theoretical methods adopted in this work to qualitatively and quantitatively investigate the ground-state properties and elementary collective excitations of a dipolar quantum gas confined in a tunable three-dimensional harmonic trap, i.e., a variational approach using a Gaussian ansatz together with a sum-rule analysis, and a complementary numerical extended Gross-Pitaevskii equation incorporating quantum fluctuations. 

\subsection{Dipolar quantum gases in lanthanide atoms with quantum fluctuations}

To characterize typical phases such as the quantum droplet state in a dipolar quantum gas, we start with the total energy functional of a three-dimensional (3D) dipolar gas trapped in an external potential incorporating the mean-field contact interaction, dipole-dipole interaction, and quantum fluctuations~\cite{baillie2016self,bisset2016ground,wachtler2016ground}, i.e.,
\begin{eqnarray}\label{eq:dipolar-Etot}
    E^\mathrm{3D}[\psi]&=&\int d{\bf r}\psi^\ast({\bf r})\left[ -\frac{\hbar^2\nabla^2}{2m}+V_\mathrm{ext}({\bf r})+\frac{1}{2}g|\psi({\bf r})|^2\nonumber \right.\\
 &&\left.+\frac{1}{2}\Phi_\mathrm{dd}({\bf r})+\frac{2}{5}\gamma_\mathrm{QF}|\psi({\bf r})|^3\right]
    \psi({\bf r}),
\end{eqnarray}
in terms of the macroscopic wave function $\psi({\bf r})$. 
$V_\mathrm{ext}({\bf r})=\frac{1}{2}m(\omega_x^2x^2+\omega_y^2y^2+\omega_z^2z^2)$ is a tunable external harmonic potential, and we assume an axial symmetry about the $z$ axis such that $\omega_x=\omega_y=\omega_\rho$ and $V_\mathrm{ext}({\bf r})=\frac{1}{2}m\omega_\rho^2\rho^2+\frac{1}{2}m\omega_z^2z^2$ with $\rho=(x^2+y^2)^{1/2}$. 
Here, we introduce the trap aspect ratio $\lambda = \omega_z/\omega_\rho$ to characterize the anisotropy of the external confinement. Specifically, $\lambda \gg 1$ corresponds to an oblate or flattened 3D confinement, whereas $\lambda \ll 1$ represents a prolate or elongated 3D confinement.
$g=4\pi\hbar^2 a_s/m$ is the 3D effective $s$-wave contact interaction strength with the $s$-wave scattering length $a_s$, and the dipole-dipole interaction energy is given by
\begin{eqnarray}
    \Phi_\mathrm{dd}({\bf r})=\int d{\bf r}'V_\mathrm{dd}({\bf r}-{\bf r}')|\psi({\bf r}')|^2,
\end{eqnarray}
with the dipole-dipole interaction $V_\mathrm{dd}({\bf r}-{\bf r}')$, which plays an important role in lanthanide atoms such as $^{168}$Er, $^{164}$Dy~\cite{lahaye2009physics,baranov2012condensed,bottcher2021new,luo2021new,guo2021new,norcia2021developments,chomaz2023dipolar}.
In ultracold atom experiments~\cite{griesmaier2005bose,lu2011strongly,aikawa2012bose}, an external magnetic field is usually applied to align all dipoles along the same direction, e.g., the $z$ axis, and the dipole-dipole interaction becomes
\begin{eqnarray}
    V_\mathrm{dd}({\bf r}-{\bf r}')=\frac{C_\mathrm{dd}}{4\pi}\frac{1-3\cos^2\theta}{r^3},
\end{eqnarray}
where $\theta$ is the angle between the relative distance vector $({\bf r}-{\bf r}')$ and the dipole orientation (i.e., the $z$ axis). Here, $C_\mathrm{dd}=\mu_0\mu_m^2$ is the coupling constant, with $\mu_0$ the vacuum magnetic permeability and $\mu_m$ the permanent magnetic dipole moment.
The lanthanide atoms usually possess large magnetic moments of several Bohr magnetons $\mu_\mathrm{B}$.
For instance, the permanent magnetic dipole moments are about $\mu_m \approx 6\mu_\mathrm{B}$, $7\mu_\mathrm{B}$, and $10\mu_\mathrm{B}$ for chromium atoms~\cite{griesmaier2005bose} (e.g, $^{52}$Cr), erbium isotopes~\cite{aikawa2012bose} (e.g., $^{166}\mathrm{Er}$ and $^{168}\mathrm{Er}$), and dysprosium isotopes~\cite{lu2011strongly} (e.g., $^{162}\mathrm{Dy}$ and $^{164}\mathrm{Dy}$), respectively.

The quantum fluctuations are characterized by the last term in Eq.~\eqref{eq:dipolar-Etot}, and the associated coefficient is given by 
$\gamma_\mathrm{QF} = \frac{32}{3}g(\frac{a_s^3}{\pi})^{1/2}\mathcal{Q}_5(\varepsilon_\mathrm{dd})$ with the auxiliary function $\mathcal{Q}_l(x)=\int_0^1du(1-x+3xu^2)^{l/2}$ obtained from the standard Bogoliubov theory~\cite{schutzhold2006mean,lima2011quantum,lima2012beyond}, that can be further simplified to 
$\gamma_\mathrm{QF}=\frac{32}{3}g(\frac{a_s^3}{\pi})^{1/2}(1+\frac{3}{2}\varepsilon_\mathrm{dd}^2)$ by expanding $\mathcal{Q}_5(\varepsilon_\mathrm{dd})=1+\frac{3}{2}\varepsilon_\mathrm{dd}^2+{\it O}(\varepsilon_\mathrm{dd}^4)$~\cite{bisset2016ground}. 
In the present work, we consistently employ $\mathcal{Q}_5(\varepsilon_\mathrm{dd})\backsimeq1+\frac{3}{2}\varepsilon_\mathrm{dd}^2$  following Ref.~\cite{bisset2016ground} across all parameter regimes considered. Within this prescription, we have neglected the small imaginary contribution from unstable low-momentum modes for $\varepsilon_\mathrm{dd}>1$. Despite its simplified form, this approximation has been quantitatively validated by experimental observations of dipolar quantum droplets~\cite{ferrier2016observation, schmitt2016self}, such as peak density, lifetime, etc., thereby providing a reliable theoretical benchmark for the parameter regimes explored in this work. The dimensionless ratio $\varepsilon_\mathrm{dd}\equiv a_\mathrm{dd}/a_s$ can quantify the relative strength of the dipole–dipole and $s$-wave interactions with the introduced dipolar scattering length $a_\mathrm{dd}=m\mu_0\mu_m^2/12\pi\hbar^2$. 
In realistic experiments~\cite{schmitt2016self,chomaz2016quantum,bottcher2019transient,chomaz2019long}, $a_\mathrm{dd}$ is intrinsic for a specific lanthanide atom, and the $s$-wave scattering length $a_s$ is tuned spanning a wide range via the Feshbach resonance technique to explore distinct phases in this dipolar gas. 

\subsection{Variational approach with a Gaussian ansatz wave function}

To qualitatively investigate the ground states in this dipolar gas and for the later utilization by the sum-rule approach, we employ a variational approach using a cylindrically symmetric Gaussian ansatz~\cite{yi2001trapped,saito2016path} for the wave function in the cylindrical coordinates $(\rho,\theta,z)$, i.e.,
\begin{equation} \label{eq:gaussianansatz}    
    \psi^\mathrm{Gauss}(\rho,\theta,z) = \sqrt{\frac{N}{\pi^{\frac{3}{2}}\sigma_\rho^2\sigma_z}} e^{-\frac{1}{2}(\frac{\rho^{2}}{\sigma_\rho^{2}}+\frac{z^{2}}{\sigma_z^{2}})},
\end{equation}
that is normalized to a total atom number $N$ and independent of the azimuthal angle $\theta$. Two variational parameters, i.e., $\sigma_\rho$ and $\sigma_z$, represent the radial and axial characteristic lengths of the atomic cloud, respectively.
Hence, by substituting this variational ansatz into the energy functional in Eq.~\eqref{eq:dipolar-Etot}, and integrating over all dimensions, the analytic expression of the total energy per particle is obtained as
\begin{eqnarray} \label{eq:EperN}
    \epsilon_{tot}[\sigma_\rho,\sigma_z]&=&\frac{1}{4}(\frac{2\hbar^2}{m\sigma_\rho^2}+\frac{\hbar^2}{m\sigma_z^2}+2m\omega_\rho^2\sigma_\rho^2+m\omega_z^2\sigma_z^2)
    \nonumber\\
    &+&\frac{N\hbar^2[a_s-a_\mathrm{dd}f(\frac{\sigma_\rho}{\sigma_z})]}{m\sqrt{2\pi}\sigma_\rho^2\sigma_z}+\frac{8N^{\frac{3}{2}}\gamma_\mathrm{QF}}{25\sqrt{10}\pi^{\frac{9}{4}}\sigma_\rho^3\sigma_z^{\frac{3}{2}}},
\end{eqnarray}
with a real function
\begin{eqnarray}\label{eq:fx}
    f(x)=\frac{1+2x^2}{1-x^2}-\frac{3x^2\mathrm{arctanh}\sqrt{1-x^2}}{(1-x^2)^{3/2}},
\end{eqnarray}
varying in a range $[-2,1]$.
Therefore, the equilibrium configuration of the system, like the ground state, can be determined by the condition of minimizing the total energy. 
In practice, we minimize the energy in Eq.~\eqref{eq:EperN} with respect to two variational parameters $\sigma_z$ and $\sigma_\rho$ by setting $\partial \epsilon_{tot}/\partial \sigma_z= 0$ and $\partial \epsilon_{tot}/\partial \sigma_\rho= 0$, i.e.,
\begin{subequations}
\begin{eqnarray} 
-\frac{\hbar^2}{2m\sigma_z^3}+\frac{1}{2}m\omega_z^2\sigma_{z}-\frac{N\hbar^2[a_s-a_\mathrm{dd}f(\frac{\sigma_{\rho}}{\sigma_{z}})]}{m\sqrt{2\pi}\sigma_{\rho}^2\sigma_{z}^2}&& \nonumber
\\
-\frac{N\hbar^2a_\mathrm{dd}}{m\sqrt{2\pi}\sigma_{\rho}^2\sigma_{z}}\frac{\partial{f(\frac{\sigma_{\rho}}{\sigma_{z}})}}{\partial{\sigma_{z}}}{-\frac{6\sqrt{\frac{2}{5}}N^{\frac{3}{2}}\gamma_\mathrm{QF}}{25\pi^{\frac{9}{4}}\sigma_{\rho}^3\sigma_{z}^{\frac{5}{2}}}}&=&0, \label{eq:minimization-z}
\\
-\frac{\hbar^2}{m\sigma_{\rho}^3}+m\omega_\rho^2
    \sigma_{\rho}-\frac{2N\hbar^2[a_s-a_\mathrm{dd}f(\frac{\sigma_{\rho}}{\sigma_{z}})]}{m\sqrt{2\pi}\sigma_{\rho}^3\sigma_{z}} && \nonumber
\\ 
-\frac{N\hbar^2a_\mathrm{dd}}{m\sqrt{2\pi}\sigma_{\rho}^2\sigma_{z}}\frac{\partial{f(\frac{\sigma_{\rho}}{\sigma_{z}})}}{\partial{\sigma_{\rho}}}{-\frac{12\sqrt{\frac{2}{5}}N^{\frac{3}{2}}\gamma_\mathrm{QF}}{25\pi^{\frac{9}{4}}\sigma_{\rho}^4\sigma_{z}^{\frac{3}{2}}}}&=&0. \label{eq:minimization-rho}
\end{eqnarray}
\end{subequations}
to obtain the stationary solutions, i.e., the extreme points $(\sigma_{\rho_0}, \sigma_{z_0})$. Here, the explicit forms of the derivatives $\partial{f(\frac{\sigma_{\rho}}{\sigma_{z}})}/\partial{\sigma_{z}}$ and $\partial{f(\frac{\sigma_{\rho}}{\sigma_{z}})}/\partial{\sigma_{\rho}}$ in the above equations are given in Appendix~\ref{app:derivatives}.

By substituting the extreme points $(\sigma_{\rho_0}, \sigma_{z_0})$ back into the variational ansatz in Eq.~\eqref{eq:gaussianansatz}, the ground-state wave function can be constructed to calculate the expectation value of associated physical quantities and to investigate the static properties of this dipolar gas. For instance, in terms of this Gaussian ansatz, we can calculate the radial and axial root-mean-square radii of the atomic cloud, i.e.,
\begin{subequations} \label{eq:rms-radii}
\begin{eqnarray}
    \sqrt{\langle \rho^2 \rangle} &=& \sqrt{\frac{\int d{\bf r} \rho^2|\psi|^2}{N}}= \sigma_{\rho},
    \\
	\sqrt{\langle z^2\rangle}&=&\sqrt{\frac{\int d{\bf r} z^2|\psi|^2}{N}}=\frac{\sigma_{z}}{\sqrt{2}},
\end{eqnarray}
\end{subequations}
which becomes $\sqrt{\langle \rho^2 \rangle} = \sigma_{\rho_0}$, $\sqrt{\langle z^2\rangle}=\sigma_{z_0}/\sqrt{2}$ for the ground state at the obtained extrema.
The ratio $\sigma_\rho/\sigma_z$ thus characterizes the geometric aspect ratio of the ground-state density shape. 
Meanwhile, the peak density of the ground state can then be calculated as
\begin{equation} \label{eq:npeak}
n_\mathrm{peak}^\mathrm{Gauss} = \frac{N}{\pi^{3/2}\sigma_{\rho_0}^2\sigma_{z_0}},  
\end{equation}
by taking the density value at the origin (i.e., $\rho=z=0$).

For specific values of parameters such as total atom number $N$, dipolar length $a_\mathrm{dd}$, and $s$-wave scattering length $a_s$, the total energy is minimized to obtain the extreme points $(\sigma_{\rho_0}, \sigma_{z_0})$, and the radial and axial root-mean-square radii in Eq.~\eqref{eq:rms-radii} as well as the peak density in Eq.~\eqref{eq:npeak} will be calculated to be compared with the numerical results introduced later.

\subsection{A sum-rule analysis of the breathing-mode frequency}

The sum-rule approach is a powerful theoretical framework used to study the linear response, ground-state characteristics, and collective excitations of quantum fluids like Bose-Einstein condensates. 
Instead of explicitly solving the complex Bogoliubov or time-dependent GPEs for the entire spectrum of excited states, this approach allows one to easily extract the frequencies of specific collective excitations, such as phonons, monopole breathing modes, and quadrupole modes, etc.

To be specific, for a chosen physical excitation operator $F$ with a specific symmetry, one can obtain rigorous upper bounds and highly accurate estimates for the low-lying excitation frequency $\omega_\mathrm{exc}$ belonging to this symmetry via~\cite{pitaevskii2016bose}
\begin{equation}\label{eq:sumrule_w}
(\hbar\omega_\mathrm{exc})^2 \le \frac{m_1(F)}{m_{-1}(F)}, \quad \text{or} (\hbar\omega_\mathrm{exc})^2  \le \frac{m_3(F)}{m_1(F)},
\end{equation}
by analyzing the first few energy-weighted moments
\begin{equation}
    m_p(F) = \hbar^{p+1}\int_{-\infty}^{\infty} \omega^p S_F(\omega) d\omega,
\end{equation}
of the dynamic structure factor $S_F(\omega)$ that can be obtained from exact commutators. Both relations in Eq.~\eqref{eq:sumrule_w} serve as upper bounds within the sum-rule approach. Thus, one can select specific geometric operators for $F$ to isolate corresponding collective modes. 
For instance, in trapped atomic gases, typical dipole, monopole (breathing) mode, and quadrupole mode can be investigated by using a linear operator $F = \sum_i r_i$, a quadratic operator $F = \sum_i r_i^2$, and an anisotropic operator $F = \sum_i (x_i^2 - y_i^2)$, respectively.

In this work, the breathing mode and the mode frequency of a dipolar quantum gas are emphasized, which have received significant attention in recent experiments, see for example Refs.~\cite{chomaz2016quantum,tanzi2019supersolid,zhen2025breaking}. We present in detail the sum-rule analysis of the axial and radial breathing modes and derive their explicit expressions within the variational approximation. 
In practice, we employ the $m_1/m_{-1}$ relation, as the moments $m_{1,-1}$ can be conveniently evaluated within the Gaussian variational approximation.
For the axial breathing mode along the magnetic polarization direction, i.e., the $z$ axis, we introduce the excitation operator
\begin{equation}
   F_z = \sum_{i=1}^{N} z_i^2, 
\end{equation}
for the axial breathing mode, and the mode excitation frequency $\omega_{\mathrm{b},z}$ can be estimated by the upper bound $\omega_{\mathrm{b},z}^2 = \frac{m_1}{m_{-1}\hbar^2}$. Here, the first-order energy-weighted moment $m_1$ can be directly calculated via the double commutator $m_1(F_z) = \frac{1}{2}\langle \psi | [[F_z, H], F_z] | \psi \rangle$, which yields straightforwardly $[[F_z, H], F_z] = \sum_{i} \frac{4\hbar^2}{m} z_i^2$ given the first-quantized Hamiltonian $H = \sum_{i=1}^{N} \left[ -\frac{\hbar^2 \nabla_i^2}{2m} + V_{\mathrm{ext}}({\bf r}_i) \right] + \frac{1}{2} \sum_{i \neq j}^{N} U({\bf r}_i - {\bf r}_j)$. 
Here $U({\bf r}_i - {\bf r}_j)$ represents two interaction potentials between any two magnetic atoms, which include the short-range $s$-wave contact interaction and the long-range dipole-dipole interaction.
Meanwhile, the inverse energy moment $m_{-1}$ is associated with the static polarizability $\chi_{zz}$ of the system, which can be calculated straightforwardly as $m_{-1}(F_z)=\frac{1}{2}\chi_{zz} = -\frac{N}{m} \frac{\partial \langle z^2 \rangle}{\partial \omega_z^2}$~\cite{pitaevskii2016bose}. 
Therefore, by taking the Gaussian variational wave function $\psi^\mathrm{Gauss}$, the axial energy-weighted moment is given as $m_{1}(F_z)=\frac{2\hbar^2}{m}\langle \sum_{i}^{N} z_i^2\rangle=\frac{\hbar^2N}{m}\sigma_{z}^2$. Meanwhile, we apply the chain rule and obtain $\frac{\partial\langle z^2\rangle}{\partial\omega_z^2}=\frac{\partial\langle z^2\rangle}{\partial\sigma_{z}}\frac{\partial\sigma_{z}}{\partial\omega_z^2}+\frac{\partial\langle z^2\rangle}{\partial\sigma_{\rho}}\frac{\partial \sigma_{\rho}}{\partial\omega_z^2}$ in $m_{-1}(F_z)$. Based on Eq.~\eqref{eq:rms-radii}, $\langle z^2 \rangle$ depends only on $\sigma_{z}$ and thus $\frac{\partial\langle z^2\rangle}{\partial\sigma_{z}}=\sigma_{z},~\frac{\partial\langle z^2\rangle}{\partial\sigma_{\rho}}=0$. After substituting back, the axial inverse energy-weighted moment becomes $m_{-1}(F_z)=-\frac{N\sigma_{z}}{m}\frac{\partial\sigma_{z}}{\partial\omega_z^2}$.
Therefore, we reach the bounded expression of the axial breathing-mode frequency as~\cite{menotti2002collective,wachtler2016ground}
\begin{equation}
    \omega_{\mathrm{b},z}^2 = \frac{m_1}{m_{-1}\hbar^2} = -\sigma_{z} \frac{\partial \omega_{z}^2}{\partial \sigma_{z}}.
\end{equation}
Finally, after performing straightforward algebraic manipulations at the energy extreme point in Eq.~\eqref{eq:minimization-z}, we obtain the analytical expression for the breathing-mode frequency
\begin{eqnarray} \label{eq:omegaB-z}
    \omega_{\mathrm{b},z}^2 &=& \frac{4\hbar^2}{m^2\sigma_{z_0}^4} + \frac{6N\hbar^2 \left[ a_s - a_\mathrm{dd} f\left(\frac{\sigma_{\rho_0}}{\sigma_{z_0}}\right) \right]}{m^2\sqrt{2\pi} \sigma_{\rho_0}^2 \sigma_{z_0}^3} + \frac{6N\hbar^2a_\mathrm{dd}\frac{\partial f\left(\frac{\sigma_{\rho_0}}{\sigma_{z_0}}\right)}{\partial \sigma_{z_0}}}{m^2\sqrt{2\pi} \sigma_{\rho_0}^2 \sigma_{z_0}^2}  \nonumber
    \\
    && -\frac{2N\hbar^2a_\mathrm{dd}}{m^2\sqrt{2\pi} \sigma_{\rho_0}^2 \sigma_{z_0}} \frac{\partial^2 f\left(\frac{\sigma_{\rho_0}}{\sigma_{z_0}}\right)}{\partial \sigma_{z_0}^2} + \frac{42\sqrt{\frac{2}{5}} N^{\frac{3}{2}} \gamma_\mathrm{QF}}{25m\pi^{\frac{9}{4}} \sigma_{\rho_0}^3 \sigma_{z_0}^{\frac{7}{2}}},
\end{eqnarray}
in terms of the obtained widths $(\sigma_{z_0}, \sigma_{\rho_0})$ and the explicit forms of the derivatives can be seen in Appendix~\ref{app:derivatives}.

Similarly, for the radial breathing mode perpendicular to the polarization direction, i.e., the transverse $xy$ plane, the excitation operator is given by
\begin{equation}
    F_\rho = \sum_{i=1}^N \rho_i^2,
\end{equation}
with the radial radius $\rho=(x^2+y^2)^{1/2}$. 
After taking again the Gaussian variational wave function $\psi^\mathrm{Gauss}$, we can evaluate the commutators and obtain the radial first-order energy-weighted moment
$m_{1}(F_\rho)={\frac{2\hbar^2}{m}\langle \sum_{i}^{N} \rho_i^2\rangle}=\frac{2\hbar^2N}{m}\sigma_{\rho}^2$. 
Similarly, we apply the chain rule $\frac{\partial\langle \rho^2\rangle}{\partial\omega_\rho^2}=\frac{\partial\langle \rho^2\rangle}{\partial\sigma_{\rho}}\frac{\partial\sigma_{\rho}}{\partial\omega_\rho^2}+\frac{\partial\langle \rho^2\rangle}{\partial\sigma_{z}}\frac{\partial \sigma_{z}}{\partial\omega_\rho^2}$ in $m_{-1}(F_\rho)$ and use $\frac{\partial\langle \rho^2\rangle}{\partial\sigma_{\rho}}=2\sigma_{\rho},~\frac{\partial\langle \rho^2\rangle}{\partial\sigma_{z}}=0$ based on Eq.~\eqref{eq:rms-radii}. As a result, the radial inverse energy moment becomes $m_{-1}(F_\rho)=-\frac{2N\sigma_{\rho}}{m}\frac{\partial \sigma_{\rho}}{\partial \omega_\rho^2}$. Within the sum-rule approach, the frequency of the radial breathing mode is bounded as
\begin{equation}
    \omega_{\mathrm{b},\rho}^2= \frac{m_1}{m_{-1}\hbar^2} =-\sigma_{\rho}\frac{\partial \omega_\rho^2}{\partial \sigma_{\rho}}.
\end{equation}
After performing some straightforward algebra at the energy extreme point in Eq.~\eqref{eq:minimization-rho}, we derive the analytical expression for the radial breathing-mode frequency
\begin{eqnarray} \label{eq:omegaB-rho}
   \omega_{\mathrm{b},\rho}^2 &=& \frac{4\hbar^2}{m^2\sigma_{\rho_0}^4} + \frac{8N\hbar^2 \left[ a_s - a_\mathrm{dd} f\left(\frac{\sigma_{\rho_0}}{\sigma_{z_0}}\right) \right]}{m^2\sqrt{2\pi} \sigma_{\rho_0}^4 \sigma_{z_0}} + \frac{5N\hbar^2a_\mathrm{dd}\frac{\partial f\left(\frac{\sigma_{\rho_0}}{\sigma_{z_0}}\right)}{\partial \sigma_{\rho_0}}}{m^2\sqrt{2\pi} \sigma_{\rho_0}^3 \sigma_{z_0}}  \nonumber
   \\
   && -\frac{N\hbar^2a_\mathrm{dd}}{m^2\sqrt{2\pi} \sigma_{\rho_0}^2 \sigma_{z_0}} \frac{\partial^2 f\left(\frac{\sigma_{\rho_0}}{\sigma_{z_0}}\right)}{\partial \sigma_{\rho_0}^2} + \frac{12\sqrt{\frac{2}{5}} N^{\frac{3}{2}} \gamma_\mathrm{QF}}{5m\pi^{\frac{9}{4}} \sigma_{\rho_0}^5 \sigma_{z_0}^{\frac{3}{2}}},
\end{eqnarray}
Note that, the subscript $“0”$ in the derivatives of the function $f(x)$ in Eqs.~\eqref{eq:omegaB-z} and ~\eqref{eq:omegaB-rho} denotes the evaluation by substituting the extremum $(\sigma_{\rho_0},\sigma_{z_0})$ into the derivatives that are explicitly given in Appendix~\ref{app:derivatives}.

The analytic expressions of the axial breathing-mode frequency $\omega_{\mathrm{b},z}$ in Eq.~\eqref{eq:omegaB-z} and the radial breathing-mode frequency  $\omega_{\mathrm{b},\rho}$ in Eq.~\eqref{eq:omegaB-rho} are our key results in this work. For specific values of parameters such as total atom number $N$, dipolar length $a_\mathrm{dd}$, $s$-wave scattering length $a_s$, we calculate these mode frequencies at the extreme points, and make a qualitative and quantitative comparison with experimental measurements and the numerical results introduced in the next section.

Note that, this sum-rule approach within the variational approximation successfully captures the overall trends of physical quantities like the peak density, atomic sizes, and the breathing-mode frequencies, though its quantitative accuracy is influenced by certain factors such as trap geometry. The underlying limitation stems from the decoupled nature of our variational ansatz for the wave function, in which the radial and axial degrees of freedom are treated independently. While the equilibrium widths $\sigma_\rho$ and $\sigma_z$ are determined self-consistently from the coupled ground-state energy, the variational wave function itself lacks explicit radial-axial coupling. As a result, when the harmonic confinement is strongly anisotropic and the two directions become significantly coupled, the present variational framework can no longer accurately describe quantities such as the peak density and the excitation spectrum. This limitation is particularly pronounced for radial breathing modes, where the coupling effects are most relevant.
Nevertheless, within the parameter regimes considered in this work, the method provides reliable estimates for the lowest compressional modes and serves as a useful benchmark for the numerical calculations.

\subsection{The extended Gross–Pitaevskii equation} \label{sec:extendedGPE}

To validate the accuracy and reliability of the analytic results about the cloud radii, peak density, and breathing-mode frequency obtained from the variational approximation and the sum-rule approach, we further introduce the extended GPE incorporating quantum fluctuations for this dipolar gas. In the weak-interaction and dilute limits, this extended GPE can well describe the static and dynamic properties of this dipolar quantum gas under the interplay of the isotropic $s$-wave contact interaction, the anisotropic dipole-dipole interaction, and the isotropic LHY term induced by quantum fluctuations~\cite{wachtler2016quantum,baillie2016self,bisset2016ground,wachtler2016ground}.

This time-dependent extended GPE can be derived from Heisenberg equations of motion or from the variations of the total energy in Eq.~\eqref{eq:dipolar-Etot} and takes the form of 
\begin{eqnarray} \label{eq:eGPE}
i\hbar \frac{\partial \psi({\bf r}, t)}{\partial t} 
&=& \biggr[ -\frac{\hbar^2 \nabla^2}{2m} + V_{\mathrm{ext}}({\bf r}) 
      + g|\psi({\bf r}, t)|^2  \\\nonumber
&& + \Phi_{\mathrm{dd}}({\bf r}, t) 
      + \gamma_{\mathrm{QF}} |\psi({\bf r}, t)|^3 \biggr] 
      \psi({\bf r}, t),
\end{eqnarray}
with the LHY term included to investigate the potential quantum droplet state in this dipolar gas~\cite{wachtler2016quantum,baillie2016self,bisset2016ground,wachtler2016ground,schmitt2016self,chomaz2016quantum,bottcher2019dilute,tanzi2019observation,tanzi2019supersolid}

First, after applying the separation of variables to the time-dependent wave function, i.e., $\psi({\bf r}, t) = \phi({\bf r}) e^{-i\mu t/\hbar}$ with the chemical potential $\mu$, and substituting into Eq.~\eqref{eq:eGPE}, the static extended GPE is obtained and one may numerically solve it to get the ground-state wave function using the imaginary-time propagation method. With this wave function, we can further calculate the radii (i.e., $\sqrt{\langle \rho^2 \rangle}=\sqrt{\int d{\bf r} \rho^2|\phi|^2/N}$, $\sqrt{\langle z^2 \rangle}=\sqrt{\int d{\bf r} z^2|\phi|^2/N}$), and the peak density to study the associated static properties of the ground state, and compare with the previous results in Eqs.~\eqref{eq:rms-radii} and ~\eqref{eq:npeak} obtained from the variational approach.

To investigate collective excitation modes in this weakly interacting dipolar gas and quantitatively examine the analytic prediction from the variational approximation and sum-rule analysis, we design specific weak perturbations to excite the specific collective modes as done in our previous works~\cite{fei2024collective,zheng2025collective,xiao2026one} and in recent dipolar experiments~\cite{chomaz2016quantum,ferrier2018scissors,tanzi2019supersolid,zhen2025breaking}.
Here, for the concerned monopole (breathing) mode, we apply a slight perturbation to the external 3D harmonic potential $V_\mathrm{ext}({\bf r})$. 
To be specific, the radial breathing mode can be excited by slightly modifying the radial trapping frequency $\omega_\rho$ of the harmonic potential with a small parameter $\chi$, i.e., 
\begin{equation}
    V_\mathrm{ext}'({\bf r})=\frac{1}{2}m\left[(1+\chi)^2\omega_\rho^2\rho^2+\omega_z^2z^2\right],
\end{equation} 
and then the system evolves with the time-dependent extended GPE~\eqref{eq:eGPE}. Here, $\chi$ is typically on the order of $10^{-3}$ to $10^{-2}$. Such a perturbation can induce the system to exhibit breathing-like oscillations in the radial plane, namely, expansion and contraction. With the calculated time-dependent wave function $\psi({\bf r}, t)$, the oscillation frequency $\omega_{\mathrm{b},\rho}$ of the radial breathing mode can be straightforwardly extracted by analyzing the temporal oscillating observable $\langle\rho^2(t)\rangle=\int d{\bf r} \rho^2|\psi({\bf r}, t)|^2/N$ and its Fourier transform.
Similarly, we excite the axial breathing mode by perturbing the harmonic potential
\begin{equation}
    V_\mathrm{ext}'({\bf r})=\frac{1}{2}m\left[\omega_\rho^2\rho^2+(1+\chi)^2\omega_z^2z^2\right],
\end{equation} 
with the axial trapping frequency $\omega_z$ slightly modified, and evolve the system with the new external trapping potential.
Thus, the mode frequency $\omega_{\mathrm{b},z}$ can be extracted numerically by analyzing the oscillating observable $\langle z^2(t)\rangle=\int d{\bf r} z^2|\psi({\bf r}, t)|^2/N$ and its Fourier transform.

Eventually, we first calculate numerically the radii $\sqrt{\langle \rho^2 \rangle}$, $\sqrt{\langle z^2 \rangle}$, and the peak density using the ground-state wave function obtained from the static extended GPE, and make a quantitative comparison with the analytic predictions in Eqs.~\eqref{eq:rms-radii} and ~\eqref{eq:npeak} obtained from the variational approach. Moreover, we slightly modify the external potential to $V_\mathrm{ext}'({\bf r})$ and evolve the system using Eq.~\eqref{eq:eGPE}. The radial and axial breathing-mode frequencies can be extracted numerically by a Fourier analysis of the 
temporal observables $\langle \rho^2(t)\rangle$ and $\langle z^2(t)\rangle$. The numerical results of  $\omega_{\mathrm{b},\rho}$ and $\omega_{\mathrm{b},z}$ will be further compared with those from a sum-rule analysis in Eqs.~\eqref{eq:omegaB-rho} and ~\eqref{eq:omegaB-z} and the experimental measurements in Refs.~\cite{chomaz2016quantum,tanzi2019supersolid}.

Note that we have validated our perturbation scheme in the numerical simulations of the time-dependent extended GPE by explicitly computing the resulting breathing-mode frequencies and quantitatively comparing them with the full spectra obtained from the Bogoliubov–de Gennes equations in Refs.~\cite{baillie2017collective,blakie2023axial}, as detailed in Appendix~\ref{app:GPEvsBdG}.

\section{RESULTS AND DISCUSSIONS}\label{sec:results}

In this section, we systematically investigate the ground-state phase diagrams, static properties, the role of confinement anisotropy, and emphasize the breathing-mode frequency of dipolar quantum gases by quantitatively comparing with experimental measurements.

In the numerical calculations, unless otherwise specified, we consider a dipolar gas of $^{164}$Dy atoms with the total atom number $N$ ranging from $10^2$ to $10^4$ and a relatively large magnetic dipole moment of $\mu_m = 9.93\mu_\mathrm{B}$, yielding a dipolar length of $a_{\text{dd}} \approx 130.8a_0$~\cite{kadau2016observing,ferrier2016observation}. The radial trapping frequency of the external confinement is fixed at $\omega_\rho = 2\pi \times 45$ Hz, while the axial one is tunable, leading to a variable trap aspect ratio $\lambda\equiv\omega_z/\omega_\rho$ varying in the range of $[0.1,10]$. The $s$-wave scattering length can be feasibly tuned in realistic experiments, which spans over a range of $[40,130]a_0$ in this work.
In the extended GPE simulations, the three-dimensional system size takes $L_x=L_y=10l_0, L_z=20l_0$ with grid points $2^7\times2^7\times2^8$, and the time step is about $10^{-4}t_0\sim10^{-3}t_0$ for both imaginary- and real-time evolution to calculate static and dynamical wave functions, respectively. Here, we have adopted $l_0\equiv\sqrt{\hbar/(m\omega_\rho)}$, $t_0\equiv\omega_\rho^{-1}$ as the units of length and time.

\begin{figure}[t]
\centering
\includegraphics[width=0.48\textwidth]  
{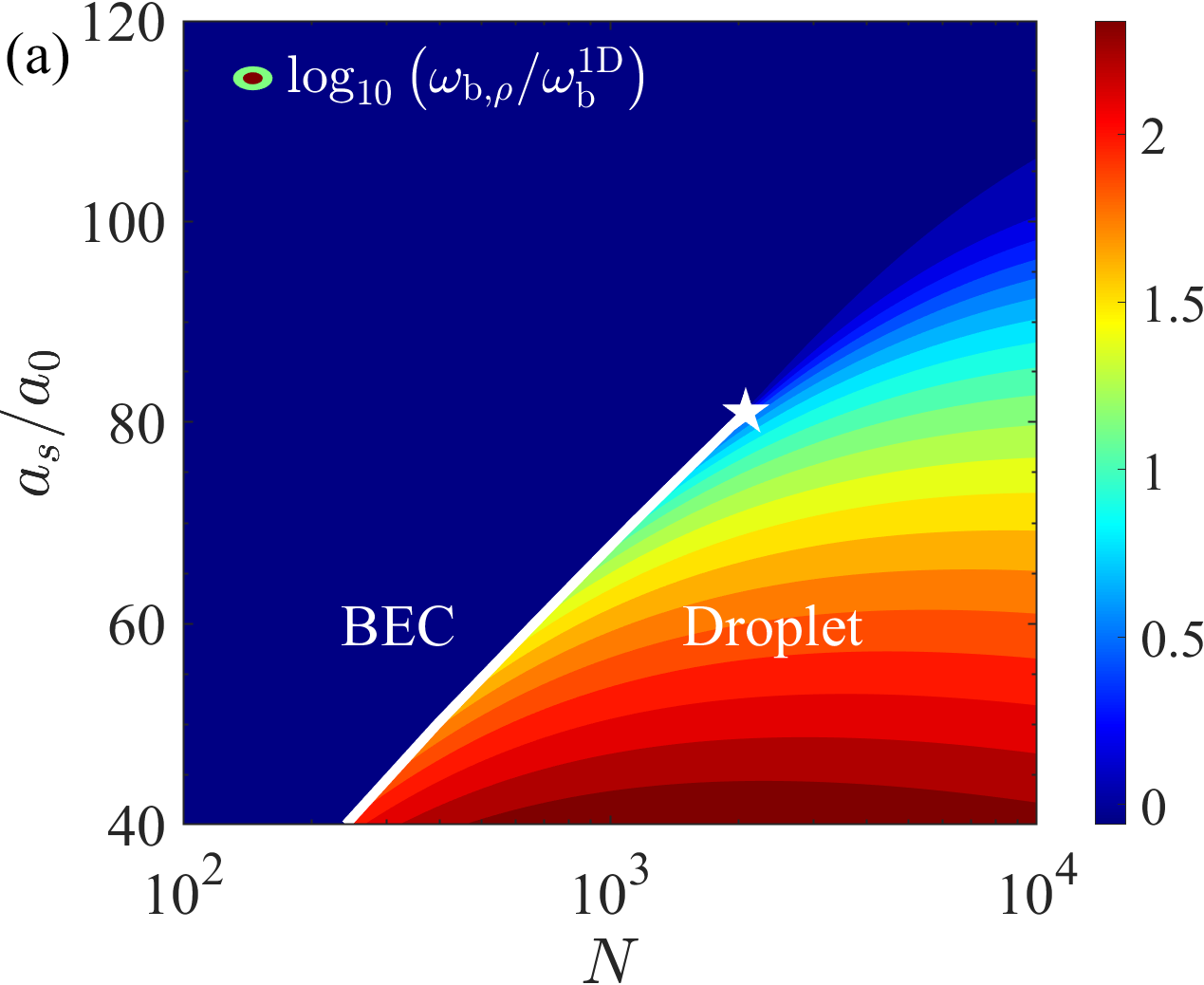}
\\
\includegraphics[width=0.48\textwidth]  {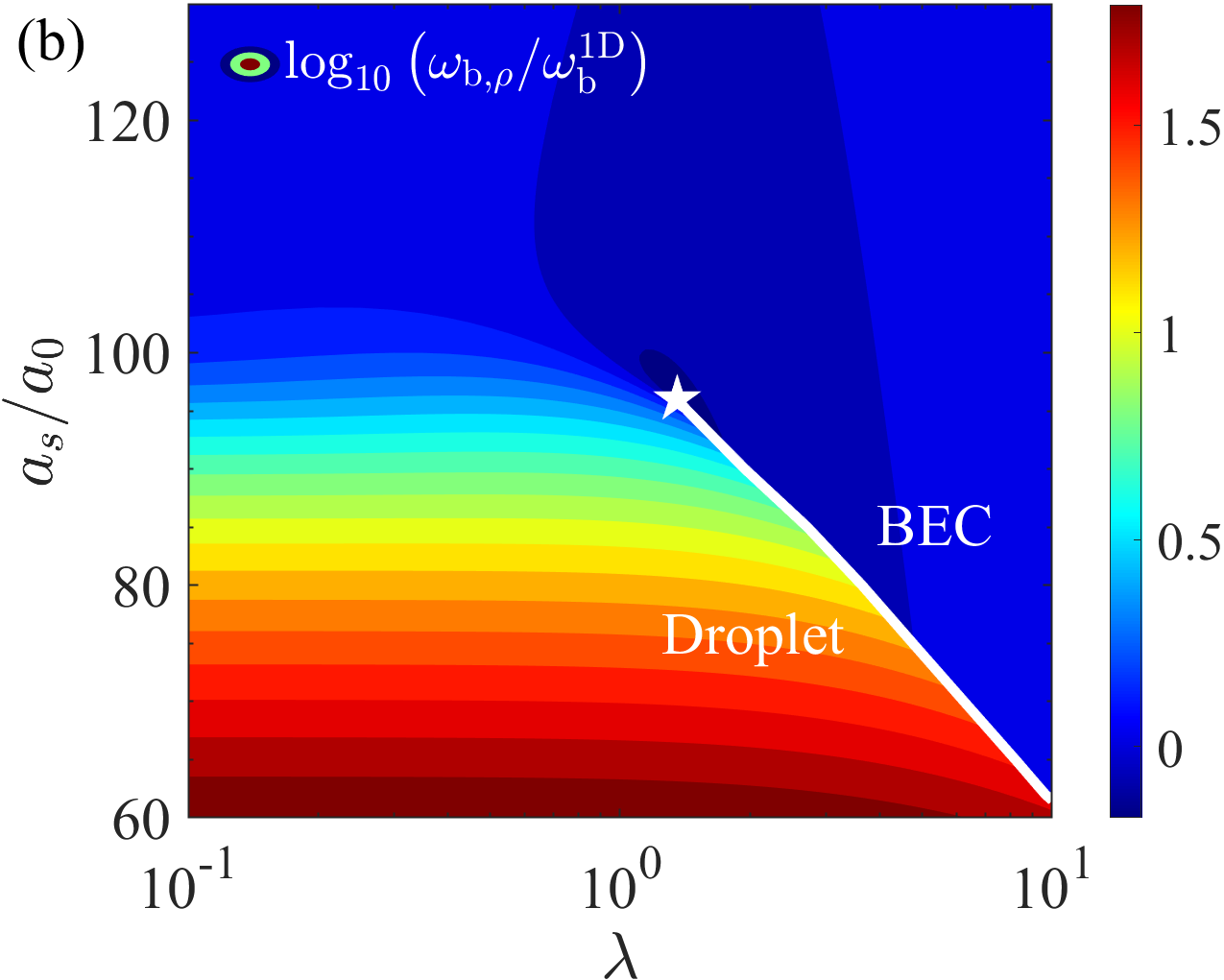}
\caption{\justifying
Phase diagrams in the parameter spaces of the $s$-wave scattering length $a_s$, the total atom number $N$, and the trap aspect ratio $\lambda\equiv\omega_z/\omega_{\rho}$, illustrated by the scaled radial breathing-mode frequency $\log_{10}\left(\omega_{\mathrm{b},\rho}/\omega_\mathrm{b}^{1\mathrm{D}}\right)$ obtained via Eq.~\eqref{eq:omegaB-rho}. $\omega_\mathrm{b}^{1\mathrm{D}}=\sqrt{3}\omega_\mathrm{trap}$ is the breathing-mode frequency for a conventional one-dimensional weakly interacting Bose gas. The white curves indicate the transition lines from the BEC to the droplet state, and the pentagrams denote the critical points at about $(N_{cp}=2.08\times10^3$, $a_{s,cp}=81a_0)$ and $(\lambda_{cp}=1.4$, $a_{s,cp}=96a_0)$. Here, we have taken $^{164}$Dy atoms confined in a cylindrically symmetric harmonic trap with a fixed radial frequency $\omega_\rho= 2\pi\times 45$ Hz for (a) $\omega_z = 2\pi\times 22.5$ Hz, and (b) a tunable axial frequency $\omega_z = \lambda\omega_\rho$ with $N=10^4$, respectively.} 
\label{fig1:phasediagrams}
\end{figure}

\subsection{Phase diagrams}

The typical phase diagrams in $^{164}$Dy atoms are presented in the parameter spaces of the $s$-wave scattering length $a_s$, the total atom number $N$, and the trap aspect ratio $\lambda\equiv\omega_z/\omega_{\rho}$ in Fig.~\ref{fig1:phasediagrams}. The gradient colors illustrate the scaled radial breathing-mode frequency $\log_{10}\left(\omega_{\mathrm{b},\rho}/\omega_\mathrm{b}^{1\mathrm{D}}\right)$ obtained via the sum-rule prediction in Eq.~\eqref{eq:omegaB-rho}, with $\omega_\mathrm{b}^{1\mathrm{D}}=\sqrt{3}\omega_\mathrm{trap}$ being the breathing-mode frequency of a conventional one-dimensional weakly interacting Bose gas.
Two typical phases, i.e., the high-density quantum droplet state and the low-density BEC state, can be identified in the parameter spaces~\footnote{We note that neither the supersolid-like phase nor the droplet-array phase is accessed in the configurations considered here.}.
For the droplets, the total energy and the chemical potential are negative relative to the free-atom threshold, confirming the self-bound nature. In contrast, two quantities become positive in the BEC-like regime, and the system relies on the dominant confinement to be bounded. The white curves indicate the phase transition lines between two phases, which are determined by the energy degeneracy point of the two individual states. Within this variational approach, these two states manifest clearly as two local minima on the energy surface in the parameter spaces.
The phase transition occurs when the energies of the two states become degenerate. On either side of the transition line, the droplet and BEC states each become the global minimum, and the system accordingly exhibits two distinct states. In addition, the white pentagrams mark the critical points at which the two states merge into a single solution, and the energy difference between the droplet and BEC states diminishes. Note that the phase diagram remains unchanged if the axial breathing-mode frequency $\omega_{\mathrm{b},z}$ is plotted instead.

In Fig.~\ref{fig1:phasediagrams} (a), the scaled radial breathing-mode frequency is shown for $^{164}$Dy atoms confined in a cylindrically symmetric harmonic trap with frequencies $(\omega_\rho,\omega_z)= 2\pi\times (45,22.5)$ Hz, i.e., an aspect ratio $\lambda\equiv\omega_z/\omega_{\rho}=0.5$. 
At relatively small $N$, the kinetic energy (quantum pressure) and the external confinement play a key role, and the system favors the BEC state. As a result, the breathing-mode frequency of the system approaches that of the conventional weakly interacting Bose gas, i.e., $\log_{10}\left(\omega_{\mathrm{b},\rho}/\omega_\mathrm{b}^{1\mathrm{D}}\right)\approx0$, as shown by the deep blue regime. As $N$ increases, the positive mean-field interaction and the quantum fluctuation term interplay with the negative dipolar interaction energy, supporting the quantum droplet state. At a certain atom number, as $a_s$ decreases, the negative dipolar interaction energy becomes dominant over the mean-field interactions, and the system favors the self-trapping quantum droplet state via either phase transitions or crossovers. In contrast to that of the BEC state, the breathing-mode frequency of droplets is significantly enhanced, as shown by the gradient-colored regimes. Note that, in the absence of the external confinement, the system can host only the quantum droplet state and will collapse in the small $a_s$ and $N$ regimes owing to the negative dipolar interactions.

Similarly, in Fig.~\ref{fig1:phasediagrams} (b), the scaled radial breathing-mode frequency is shown for $10^4$ $^{164}$Dy atoms confined in a cylindrically symmetric harmonic trap with a fixed radial frequency $\omega_\rho= 2\pi\times 45$ Hz and a tunable axial frequency $\omega_z=\lambda\omega_{\rho}$. In general, a relatively large scattering length leads to a dominant repulsive mean-field interaction energy, and the system favors the BEC state, see the deep blue regimes where the breathing-mode frequency approaches that of the conventional weakly interacting Bose gases. In contrast, at relatively small scattering lengths, the interplay of the dipolar interaction, mean-field interaction and quantum fluctuations stabilizes the self-bound quantum droplet state in the system with the significantly enhanced breathing-mode frequency denoted by the gradient-colored regimes. At a relatively small trap aspect ratio $\lambda$, i.e., elongated 3D confinement, the system experiences a droplet-BEC crossover as $a_s$ decreases. This is because the attractive dipolar interaction elongates the droplet in the axial direction over the radial plane regardless of the external confinement. As a consequence, small $\lambda$ or the elongated confinement can not significantly affect the droplet shape, and the crossovers become independent of $\lambda$.
In sharp contrast, at sufficiently large $\lambda$, the role of the flattened 3D confinement becomes crucial, and the elongated droplet is significantly suppressed in the axial direction. Consequently, the system gradually favors the emergence of the BEC state under the dominant external confinement as $\lambda$ increases. In addition, under this flattened confinement, the system experiences phase transitions from the BEC state to the droplet state as $a_s$ decreases, in contrast to the crossovers for the elongated confinement.

\begin{figure}[t]
\centering
\includegraphics[width=0.48\textwidth]{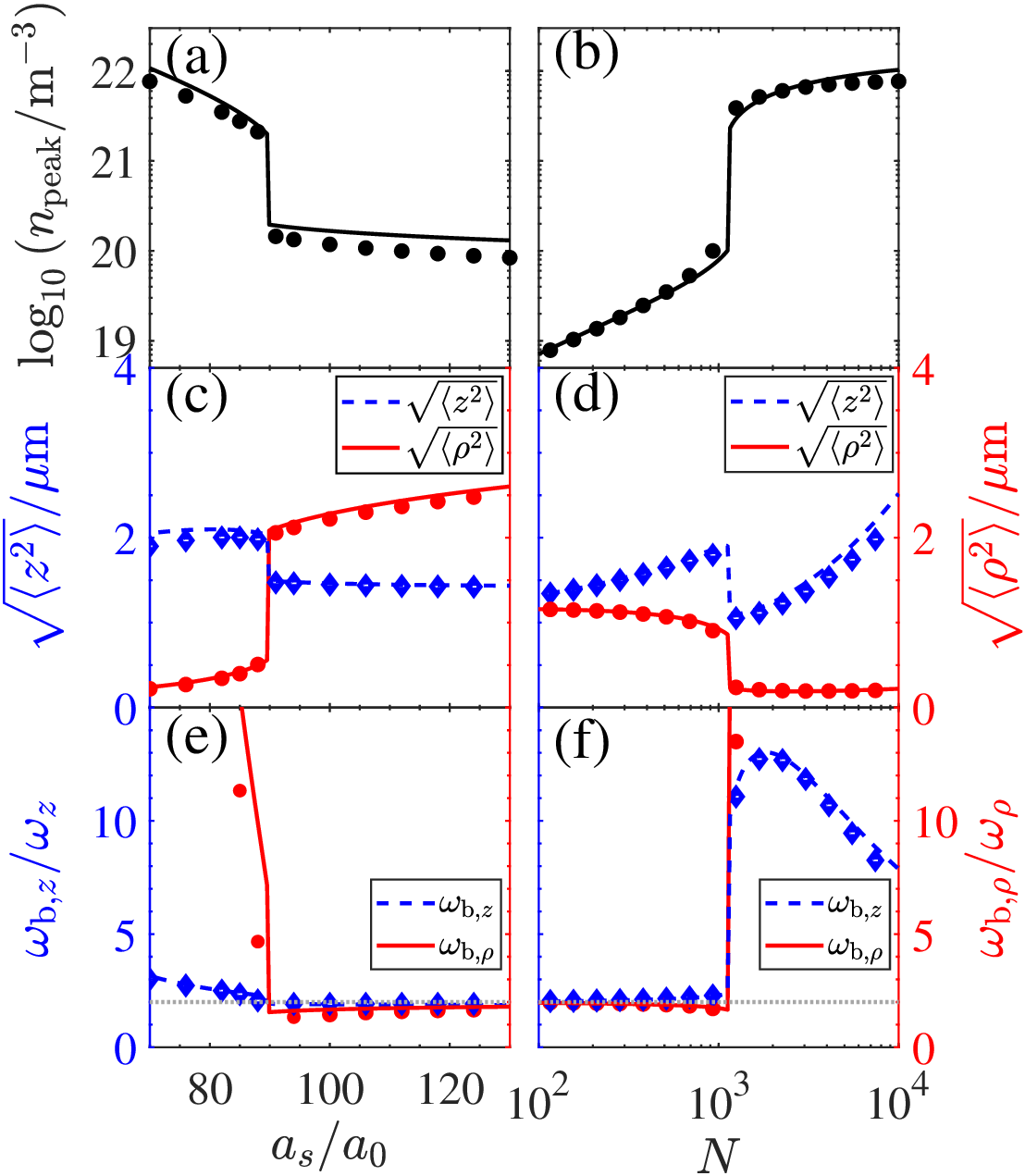}
\caption{\justifying
Phase transitions in a trapped dipolar quantum gas for $^{164}$Dy atoms, as functions of the $s$-wave scattering length $a_s$ and the particle number $N$. Panels (a) and (b) present the peak density $n_\mathrm{peak}$. Panels (c) and (d) illustrate the radial width $\sqrt{\langle\rho^2\rangle}$ (red) and the $z$-axis size $\sqrt{\langle z^2\rangle}$ (blue). Panels (e) and (f) display the radial and axial breathing-mode frequencies $\omega_{\mathrm{b},\rho}$ (red) and $\omega_{\mathrm{b},z}$ (blue).
In all plots, the lines denote the variational results using a Gaussian ansatz combined with the sum-rule approach, while the symbols indicate the numerical comparisons obtained by solving the extended GPE.
The horizontal gray dotted lines in panels (e) and (f) indicate the analytic breathing-mode frequency $\omega_\mathrm{b}^{2\mathrm{D}}=2\omega_\mathrm{trap}$ for a conventional two-dimensional weakly interacting Bose gas~\cite{ho1999quasi}.
Here, we have considered $N=10^4$ atoms with an aspect ratio $\lambda \equiv \omega_z/\omega_\rho = 2$ for the left column, and a fixed scattering length $a_s = 70a_0$ with $\lambda = 0.5$ for the right column, respectively.}
\label{fig2:phasetransitions}
\end{figure}

\subsection{General properties}
In this section, we present comprehensively the phase transitions and crossovers depicted in the phase diagrams described above by selecting specific values of the scattering length and the atom number, and make a quantitative comparison of the breathing-mode frequency with the numerical results by solving the extended GPE and the measurements in the experiments of $^{166}$Er and $^{162}$Dy atoms.

\subsubsection{Phase transitions}

In Fig.~\ref{fig2:phasetransitions}, two phase transitions between the BEC and droplet states are illustrated by investigating the peak density, atomic sizes, and the breathing-mode frequencies of a trapped $^{164}$Dy dipolar quantum gas within the variational approach. In the left column, the properties of $N=10^4$ atoms in a trap with $\lambda \equiv \omega_z/\omega_\rho = 2$ are studied as a function of the tunable scattering length $a_s$, see also the right part of Fig.~\ref{fig1:phasediagrams} (b). As $a_s$ increases, the system experiences a first-order phase transition at about $a_{s,cp}\approx 90a_0$ from the high-density quantum droplet state to the low-density BEC state with a discontinuity existing in the peak density $n_\mathrm{peak}$, as shown by the solid line in Fig.~\ref{fig2:phasetransitions} (a). Meanwhile, the radial width $\sqrt{\langle\rho^2\rangle}$ and the $z$-axis size $\sqrt{\langle z^2\rangle}$ also experience a discontinuous jump along with the phase transition, see the red-solid and blue-dashed curves in Fig.~\ref{fig2:phasetransitions} (c), respectively. In the droplet regime, the $z$-axis size is much more elongated than the radial width owing to the attractive dipolar interactions along the $z$ direction. In the BEC regime, the dominant mean-field interaction is isotropic and notably reduces the gap between these two sizes. The $z$-axis size is strongly squeezed under this trap aspect ratio $\lambda=2$ and becomes smaller than the radial width. In Fig.~\ref{fig2:phasetransitions} (e), the associated radial and axial breathing-mode frequencies $\omega_{\mathrm{b},\rho}$ and $\omega_{\mathrm{b},z}$ are shown by the red and blue curves, respectively, that are obtained from Eqs.~\eqref{eq:omegaB-rho} and~\eqref{eq:omegaB-z}. 
\begin{figure}[t]
\centering
\includegraphics[width=0.48\textwidth]{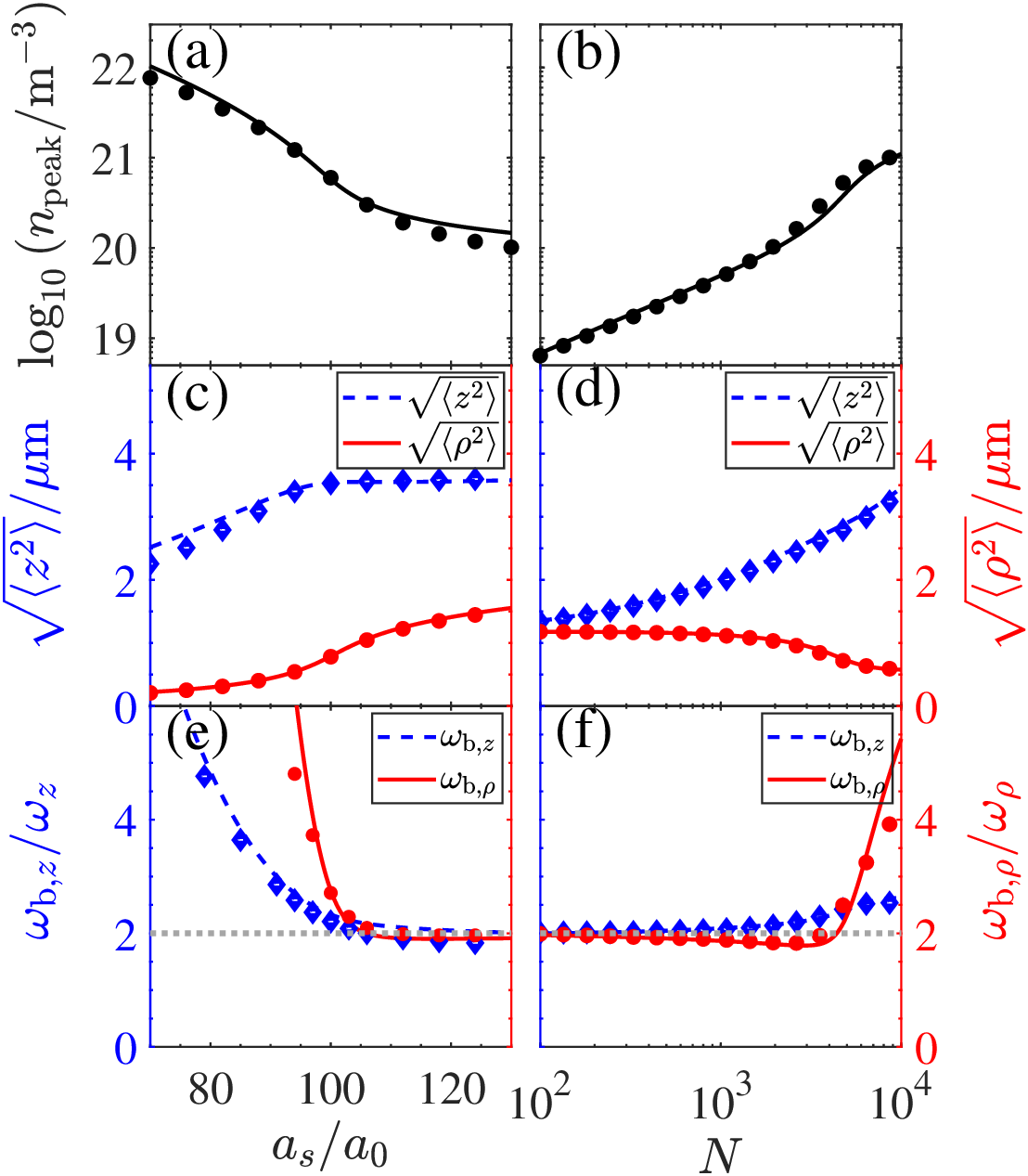}
\caption{\justifying
Crossovers in a trapped dipolar quantum gas for $^{164}$Dy atoms, as functions of the $s$-wave scattering length $a_s$ and the particle number $N$. Here, we have considered $N=10^4$ atoms with an aspect ratio $\lambda \equiv \omega_z/\omega_\rho = 0.5$ for the left column, and a fixed scattering length $a_s = 95a_0$ with $\lambda = 0.5$ for the right column, respectively. The other properties of the figure are the same as Fig.~\ref{fig2:phasetransitions}.} 
\label{fig3:crossovers}
\end{figure}
In the BEC regime with relatively large scattering lengths, the mode frequencies approach that of conventional two-dimensional weakly interacting Bose gases~\cite{ho1999quasi}, i.e., $\omega_\mathrm{b}^{2\mathrm{D}}=2\omega_\mathrm{trap}$, denoted by the dotted horizontal line. In contrast, the mode frequencies in the droplet regime are significantly enhanced owing to the high density and tend to increase monotonically as $a_s$ decreases. At the high densities characteristic of droplets, the repulsive LHY energy density scaling as $n^{5/2}$ grows faster than the linear mean-field and dipolar energy densities scaling as $n^2$. Thus, compression dramatically enhances the repulsive restoring force from the LHY term, substantially raising the breathing-mode frequencies. This restoring force is stronger in the tightly confined radial direction, while the axial direction releases energy more readily through elongation and thus exhibits a lower mode frequency.

Similarly, we fix $a_s = 70a_0$ at $\lambda =\omega_z/\omega_\rho = 0.5$ and present the corresponding peak density, atomic sizes, and breathing-mode frequencies in the right column of Fig.~\ref{fig2:phasetransitions}, see also the lower part of Fig.~\ref{fig1:phasediagrams} (a). The phase transition can be seen with the same discontinuities appearing in these quantities at about $N_{c}\approx 1.12\times10^3$ and the physical behaviors are the same. 
The filled symbols in all figures are numerical results obtained by solving the extended GPE~\eqref{eq:eGPE} incorporating quantum fluctuations. In general, the variational predictions denoted by lines show good agreement with the extended GPE results for both static properties and breathing-mode frequencies in both the droplet and BEC regimes.

\subsubsection{Crossovers}

We now present the crossovers between the BEC and droplet states of a trapped $^{164}$Dy dipolar quantum gas in Fig.~\ref{fig3:crossovers}, by illustrating the peak density, atomic sizes, and breathing-mode frequencies within the variational approach. 
We take a nearly elongated confinement with an aspect ratio $\lambda \equiv \omega_z/\omega_\rho = 0.5$ and consider a fixed atom number $N=10^4$ and a fixed scattering length $a_s = 95a_0$ for the left and right columns, respectively, see also the behaviors of the radial breathing-mode frequencies from the right part of Fig.~\ref{fig1:phasediagrams} (a) and the left part of Fig.~\ref{fig1:phasediagrams} (b). 
In sharp contrast to the phase transitions described in the previous section, the system hosts only a single minimum (or solution) within the variational approach, which could be either the BEC or the droplet state. As a result, the calculated peak density, atomic sizes, and breathing-mode frequencies experience a smooth and continuous crossover as the scattering length or the atom number varies. In the left column, the peak density $n_\mathrm{peak}$ of the system drops smoothly by one or two orders of magnitude as the scattering length increases, i.e., crossover from the droplet to the BEC. Meanwhile, in this elongated confinement, the $z$-axis size $\sqrt{\langle z^2\rangle}$ is always larger than the radial width $\sqrt{\langle\rho^2\rangle}$ regardless of the BEC or droplet regimes. As $a_s$ increases, both sizes increase monotonically with the size ratio $\sqrt{\langle z^2\rangle/\langle\rho^2\rangle}$ approaching 2, and the breathing-mode frequencies exhibit a monotonically decreasing behavior, which is significantly enhanced in the left droplet regime.

Similarly, in the right column, the system experiences a smooth droplet-BEC crossover as $N$ rises. At relatively small $N$, the interactions are negligible, and the trapping potential makes the two sizes relatively close to each other. Owing to the harmonic trap, the system tends to host a Gaussian-like wave function, and the breathing-mode frequency approaches the analytic one $\omega_\mathrm{b}^{2\mathrm{D}}=2\omega_\mathrm{trap}$ of a quantum harmonic oscillator. As $N$ rises, the emergent droplet is elongated in the $z$ axis, and the radial width is clearly squeezed. Meanwhile, the breathing-mode frequencies start from those of a conventional weakly interacting Bose gas in the BEC regime and then increase continuously in the droplet regime. Again, the numerical results (i.e., filled symbols) obtained from the extended GPE agree well with the variational predictions denoted by lines across the droplet–BEC crossover.
Together with Fig.~\ref{fig2:phasetransitions}, we can clearly see two distinct behaviors in this dipolar quantum gas, i.e., the discontinuous phase transitions and continuous crossovers, driven by parameters like scattering length, atom number, and geometric anisotropy of the confinement. 

\subsubsection{Breathing-mode frequency comparison: Experiments with \texorpdfstring{$^{166}$Er}{Lg} and \texorpdfstring{$^{162}$Dy}{Lg} atoms}
\begin{figure}[t]
\centering
\includegraphics[width=0.48\textwidth]{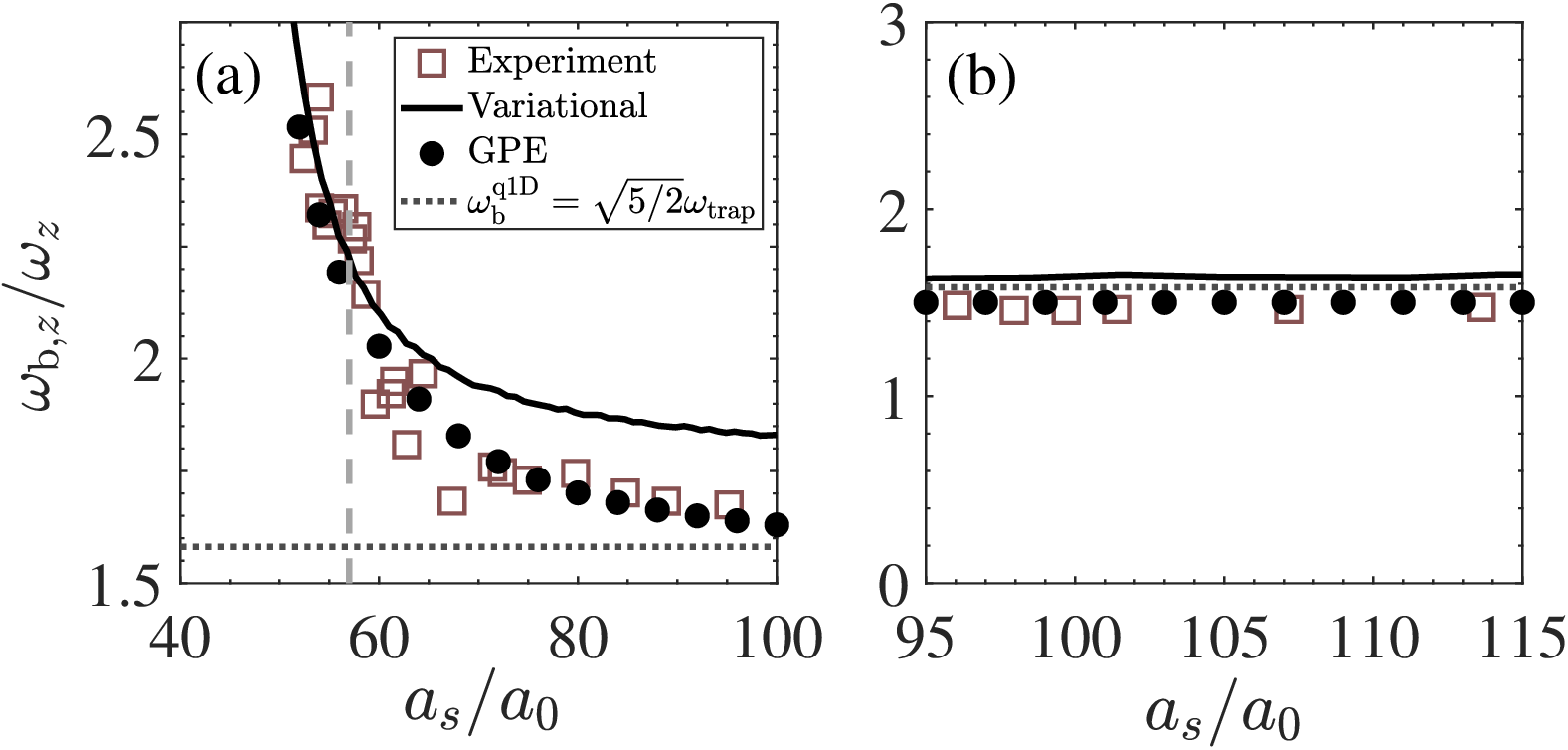}
\caption{\justifying
Comparison of breathing-mode frequency with recent experiments on dipolar quantum gases. The axial breathing-mode frequency $\omega_{\mathrm{b},z}$ is shown as a function of the $s$-wave scattering length $a_s$ for (a) $1.2\times10^5$ $^{166}$Er atoms as in Ref.~\cite{chomaz2016quantum}, and (b) $3.5\times10^4$ $^{162}$Dy atoms as in Ref.~\cite{tanzi2019supersolid}, respectively.
The hollow squares indicate the experimental results, compared with the variational results [i.e., Eq.~\eqref{eq:omegaB-z}, denoted by solid lines] using a Gaussian ansatz combined with the sum-rule approach and the numerical results (denoted by solid circles) by solving the extended Gross–Pitaevskii equation~\eqref{eq:eGPE}.
The vertical gray dashed line in panel (a) indicates the collapsing point at $a_s\approx57a_0$ in the mean-field level in Ref.~\cite{chomaz2016quantum}.
Here, the horizontal gray dotted line indicates the analytic breathing-mode frequency $\omega_\mathrm{b}^\mathrm{q1D}=\sqrt{5/2}\omega_\mathrm{trap}\approx1.58\omega_\mathrm{trap}$ of a conventional weakly interacting Bose gas trapped in an elongated 3D confinement~\cite{stringari1996collective,stringari1998dynamics}.} 
\label{fig4:comparison-exps}
\end{figure}

To validate the reliability of our theoretical approaches and results, we make a quantitative comparison of the breathing-mode frequency with two recent experiments on dipolar quantum gases in $^{166}$Er~\cite{chomaz2016quantum} and $^{162}\mathrm{Dy}$~\cite{tanzi2019supersolid} atoms. In Fig.~\ref{fig4:comparison-exps}, three values of the axial breathing-mode frequency $\omega_{\mathrm{b},z}$ are compared as a function of the $s$-wave scattering length $a_s$, which are, in sequence, the experimental measurements, the variational results from Eq.~\eqref{eq:omegaB-z} using a Gaussian ansatz combined with the sum-rule approach, and the numerical calculations by solving the extended GPE~\eqref{eq:eGPE}, denoted by hollow squares, solid lines, and solid circles, respectively.

In the first experiment~\cite{chomaz2016quantum} across a droplet-BEC crossover, about $1.2\times10^5$ $^{166}$Er atoms are trapped in an elongated confinement with harmonic trapping frequencies $(\omega_x, \omega_y, \omega_z) = 2\pi\times (156, 198, 17.2)$ Hz, giving rise to an effective trap aspect ratio $\lambda_\mathrm{eff}\equiv\omega_z/\sqrt{\omega_x\omega_y} \approx 0.097$. By transiently modulating the optical dipole trap, they measure the evolution of the axial spatial cloud width after time-of-flight (TOF) expansion using absorption imaging, and then extract the breathing-mode frequency via a damped-sine fit of the temporal width, as shown by the hollow squares in Fig.~\ref{fig4:comparison-exps}(a). 
In general, the experimental measurements agree excellently with those obtained by numerically solving the extended GPE even without considering three-body losses over the entire range of $a_s$, where the collapsing point shown by the vertical gray dashed line is predicted at $a_s\approx57a_0$ in the mean-field level~\cite{chomaz2016quantum}.
In comparison, the variational predictions using the Gaussian ansatz show good agreement with these two results only in the droplet regime (i.e., at relatively small $a_s$), while a systematic overestimation of the mode frequency is observed in the large-$a_s$ BEC regime.
At relatively large $a_s$, the dominant mean-field interaction favors the BEC, and thus the breathing-mode frequency approaches that of a quasi-one-dimensional conventional weakly interacting Bose gas~\cite{stringari1996collective,stringari1998dynamics}, i.e., $\omega_\mathrm{b}^\mathrm{q1D}=\sqrt{5/2}\omega_\mathrm{trap}\approx1.58\omega_\mathrm{trap}$. As $a_s$ decreases, the relatively stronger dipolar interaction leads the system into the droplet, and the breathing-mode frequency becomes significantly enhanced. These behaviors are consistent with those found in this work, as described in the small-$\lambda$ limit of Figs.~\ref{fig1:phasediagrams} (b) and ~\ref{fig3:crossovers} (e). 

Similarly, in the second experiment~\cite{tanzi2019supersolid}, approximately $3.5\times10^4$ $^{162}$Dy atoms are confined in an anisotropic 3D trap with $(\omega_x,\omega_y,\omega_z) = 2\pi\times (18.5, 53, 81)$ Hz, and the system is in the BEC regime over the range of $a_s\in[95,115]a_0$. Note that for relatively smaller $a_s$, the experiment enters the supersolid regime with a double- or multi-peak structure, which is beyond the scope of our single-Gaussian ansatz and thus not considered here. The associated breathing mode is excited by a rapid quench of the scattering length, and the mode frequency is ultimately extracted by analyzing the momentum-space density distribution after a long TOF expansion using the same damped-sine fit. In this range of $a_s$, the mode frequency is almost constant, independent of the scattering length. The experimental measurement and numerical result show an excellent agreement, being around $\omega_{\mathrm{b},z}\approx 1.5\omega_z$, while the variational prediction $\omega_{\mathrm{b},z}^\mathrm{var}\approx 1.63\omega_z$ is slightly higher than the other two, as shown by the solid line in Fig.~\ref{fig4:comparison-exps}(b). This is because the adopted sum-rule approach usually provides an upper bound that overestimates the frequencies of collective modes.

\subsection{Anisotropy of the external confinement}

In this section, we investigate the role of anisotropy in the external confinement on the static and dynamic properties of this dipolar quantum gas, and emphasize the elongated and flattened limits that are of great interest in cold atom experiments.

\begin{figure}[t]
\centering
\includegraphics[width=0.48\textwidth]{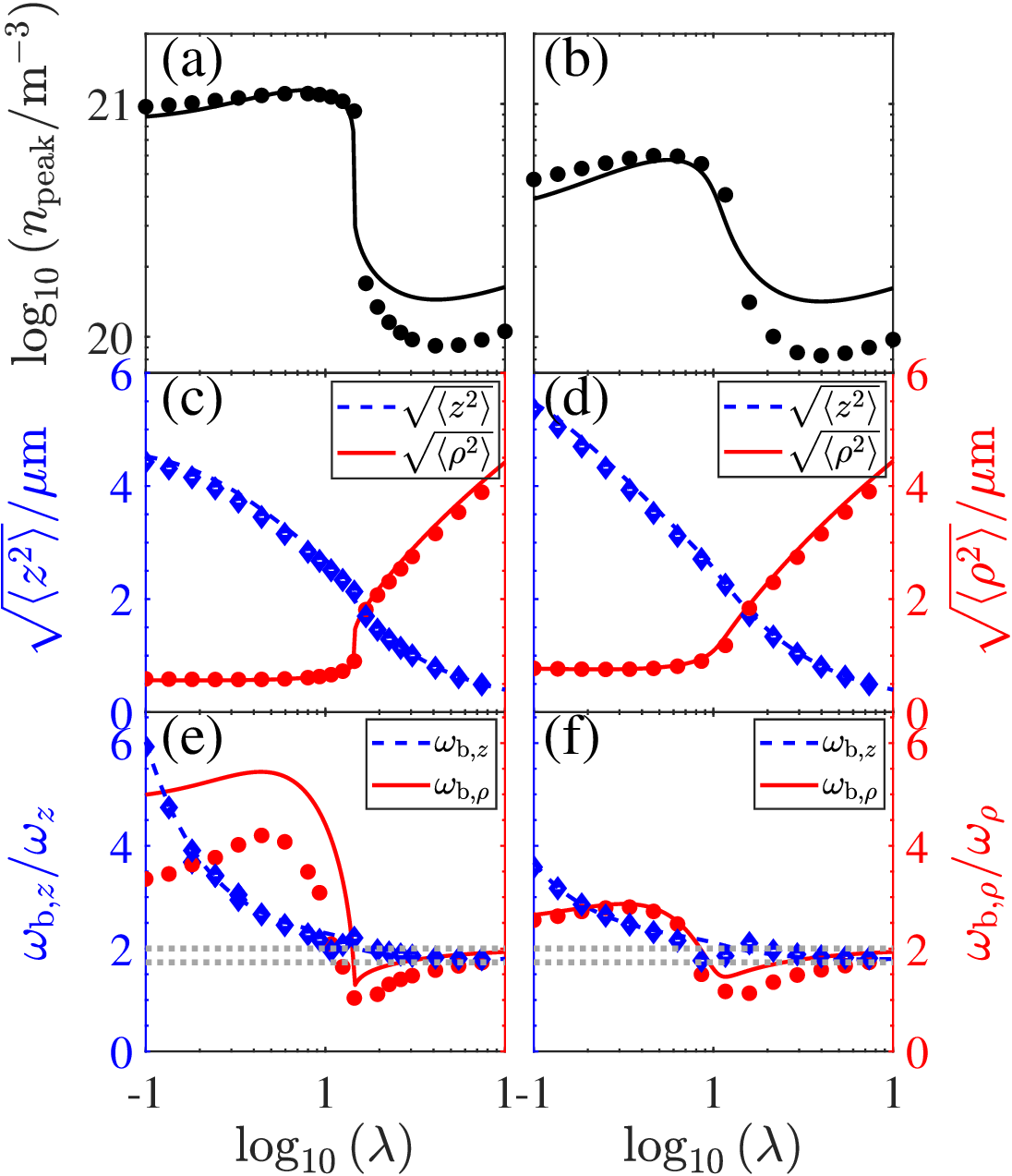}
\caption{\justifying
Phase transitions and crossovers in a trapped dipolar quantum gas for $^{164}$Dy atoms, as functions of the aspect ratio $\lambda$ of the harmonic trap. The horizontal gray dotted lines in panels (e) and (f) indicate the analytic breathing-mode frequencies $\omega_\mathrm{b}^{1\mathrm{D}}=\sqrt{3}\omega_\mathrm{trap}$ and $\omega_\mathrm{b}^{2\mathrm{D}}=2\omega_\mathrm{trap}$ for conventional one- and two-dimensional weakly interacting Bose gases~\cite{ho1999quasi,kimura2002breathing}, respectively. Here, we have considered $N=10^4$ atoms with scattering lengths $a_s=95a_0$ and $100a_0$ for the left and right columns, respectively. The other properties of the figure are the same as Fig.~\ref{fig2:phasetransitions}. 
} 
\label{fig5:anisotropy}
\end{figure}

In Fig.~\ref{fig5:anisotropy}, two distinct phase transitions and crossovers driven by a variable trap aspect ratio $\lambda\equiv\omega_z/\omega_\rho$ are depicted for $^{164}$Dy dipolar quantum gases, see also Fig.~\ref{fig1:phasediagrams} (b). We take two scattering lengths $a_s = 95a_0$ and $a_s = 100a_0$ in the left and right columns, respectively, and characterize both a phase transition and a crossover by illustrating the peak density, atomic sizes, and the breathing-mode frequencies.
In the left column, there are clearly discontinuous jumps at about $\lambda_c\approx1.4$ in all the physical quantities, revealing a trap-anisotropy-induced first-order droplet-BEC phase transition as $\lambda$ rises. At this relatively small $a_s$, the system favors the droplet state with the $z$-axis size $\sqrt{\langle z^2\rangle}$ significantly elongated, being much larger than $\sqrt{\langle \rho^2\rangle}$. As $\lambda$ increases, the confinement tends to show a flattened structure, and the $z$-axis size is strongly suppressed to become smaller than the radial width. As a consequence, the dipolar interaction energy tends to be positive, together with the repulsive mean-field interaction and LHY corrections, leading the system to a low-density BEC state. In this regime, the dominant trap potential with an increasing aspect ratio $\lambda$ leads to a continuously rising cloud size ratio $\sqrt{\langle\rho^2\rangle/\langle z^2\rangle}$. 
Meanwhile, the mode frequencies in the left droplet regime are significantly enhanced again and exhibit distinct behaviors, while they tend to approach those of conventional two-dimensional weakly interacting Bose gases as $\lambda$ increases~\cite{ho1999quasi}, i.e., $\omega_\mathrm{b}^{2\mathrm{D}}=2\omega_\mathrm{trap}$. 
In this case, the sum-rule prediction for the axial breathing-mode frequency $\omega_{\mathrm{b},z}$ is consistent with the numerical result, whereas the prediction for the radial breathing-mode frequency $\omega_{\mathrm{b},\rho}$ serves as an estimated upper bound, exceeding the numerical value.

However, in the right column of Fig.~\ref{fig5:anisotropy}, all the physical quantities experience a continuous and smooth crossover between these two states as $\lambda$ varies. Particularly, under the competition of mean-field and dipolar interactions, quantum fluctuations, and the trap potential, the state exhibits a spherical density shape with isotropic sizes $\sqrt{\langle z^2\rangle}=\sqrt{\langle \rho^2\rangle}$ at $\lambda^\mathrm{iso}\approx1.5$. Similarly, the breathing-mode frequencies in the left droplet regime are significantly enhanced, while those in the right BEC regime tend to approach that of a conventional two-dimensional weakly interacting Bose gas. 

In comparison with the numerical extended-GPE results (i.e., filled symbols), the variational predictions for atomic sizes and the axial breathing-mode frequency show quantitative agreement in both the droplet and BEC regimes. In contrast, the peak density and the radial breathing-mode frequency exhibit pronounced deviations, which become more severe as the aspect ratio $\lambda$ increases. This discrepancy arises because stronger trap anisotropy enhances the radial–axial coupling that goes beyond the decoupled variational ansatz employed here.
\begin{figure*}[t]
\centering
\includegraphics[width=0.96\textwidth]{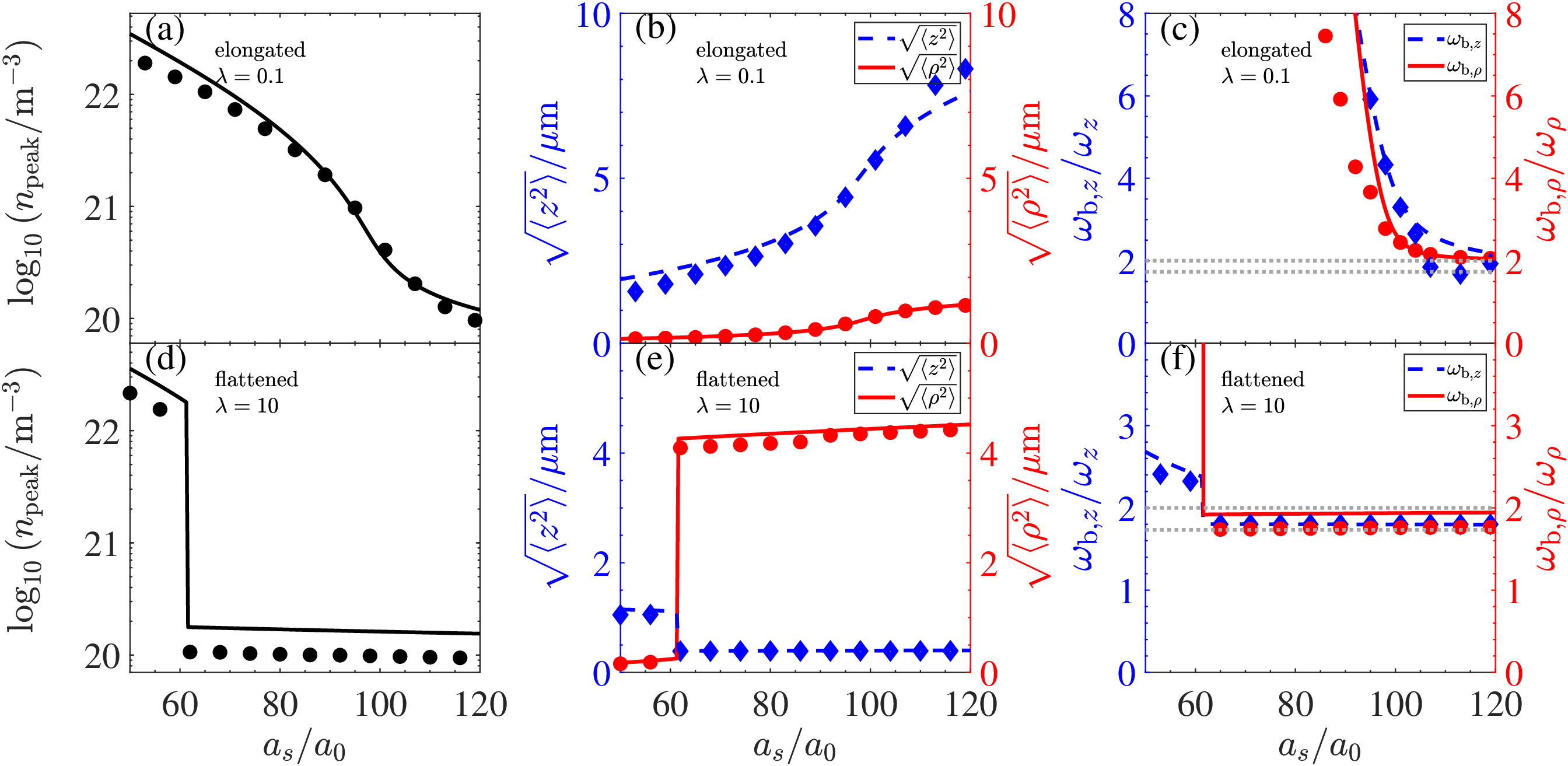}
\caption{\justifying
(a-c) Crossovers and (d-f) phase transitions of a dipolar quantum gas for $10^4$ $^{164}$Dy atoms trapped in elongated and flattened confinements with aspect ratios $\lambda \equiv \omega_z/\omega_\rho = 0.1$ and $\lambda = 10$, respectively, as functions of the tunable scattering length $a_s$. From left to right, the peak density $n_\mathrm{peak}$, the radial width $\sqrt{\langle\rho^2\rangle}$ (red) and the $z$-axis size $\sqrt{\langle z^2\rangle}$ (blue), the radial and axial breathing-mode frequencies $\omega_{\mathrm{b},\rho}$ (red) and $\omega_{\mathrm{b},z}$ (blue) are illustrated in sequence.
The other properties of the figure are the same as Fig.~\ref{fig2:phasetransitions}. }
\label{fig6:quasi1D2D}
\end{figure*}

\subsubsection{Elongated three-dimensional confinement}

By taking a typical elongated 3D confinement, i.e., at a relatively small aspect ratio $\lambda = 0.1$, the peak density, atomic sizes, and breathing-mode frequencies of a $^{164}$Dy dipolar quantum gas are presented as a function of the scattering length $a_s$ in the top panel of Fig.~\ref{fig6:quasi1D2D}. Under this elongated geometric confinement, the system exhibits a continuous droplet-BEC crossover as $a_s$ increases, see also the measurements in experiment~\cite{chomaz2016quantum}. In the left droplet regime, the dipolar interaction elongates the $z$-axis size and significantly suppresses the radial width. As $a_s$ increases, the mean-field energy drives the system towards the isotropic BEC state, and the effect of this elongated confinement starts to play a non-negligible role in the atomic sizes. Consequently, both sizes increase monotonically with $a_s$, and the $z$-axis size changes more significantly. Similar behavior of droplet widths at different $a_s$ can also be seen in Fig. 5(a) of the experiment in Ref.~\cite{chomaz2016quantum}. 
Meanwhile, as $a_s$ decreases, both the radial and axial breathing-mode frequencies, i.e., $\omega_{\mathrm{b},\rho}$ (red) and $\omega_{\mathrm{b},z}$ (blue), exhibit a monotonically increasing trend from the asymptotic values of the conventional BEC state to the significantly enhanced ones of the droplet state. This continuous rise in mode frequencies signifies an enhancement of the system's incompressibility, demonstrating that in the elongated geometry limit, the system can bypass the discontinuous collapse point and smoothly transition into a highly self-bound droplet state. The same behavior can also be seen in the experimental measurement~\cite{chomaz2016quantum} employing an elongated confinement and as shown in Fig.~\ref{fig4:comparison-exps} (a).

\subsubsection{Flattened three-dimensional confinement}

Furthermore, by taking a typical flattened 3D confinement, i.e., at a relatively large aspect ratio $\lambda = 10$, the peak density, atomic sizes, and breathing-mode frequencies of a $^{164}$Dy dipolar quantum gas are presented as a function of the scattering length $a_s$ in the bottom panel of Fig.~\ref{fig6:quasi1D2D}. In stark contrast to the case in an elongated confinement, the system in this flattened confinement experiences a discontinuous first-order droplet-BEC phase transition as $a_s$ increases, and the physical quantities exhibit pronounced discontinuities at the critical point $a_{s,cp}\approx62a_0$. The system hosts the high-density droplet state at relatively small $a_s$, the peak density drops by over two orders of magnitude at the critical point, and is almost unchanged in the BEC regime. 
In the droplet regime, the elongated $z$-axis size (in blue) is stabilized by the $z$-axis attractive dipolar interaction, and becomes larger than the radial width (in red), giving rise to an elongated density structure regardless of the flattened confinement. 
However, in the BEC regime, the dominant mean-field interaction leads the system to an isotropic BEC state that is clearly influenced by the external traps. As a result, this flattened confinement significantly squeezes the $z$-axis size, making it much smaller than the radial width. In this case, the two atomic sizes in both regimes are nearly constant and independent of $a_s$. Meanwhile, the radial and axial breathing-mode frequencies in the droplet regime are significantly enhanced, deviating strongly from those of a weakly interacting Bose gas as found in previous sections. In contrast, two breathing-mode frequencies $\omega_{\mathrm{b},\rho}$ and $\omega_{\mathrm{b},z}$ are independent of $a_s$ in the BEC regime and tend to approach those of conventional one- and two-dimensional weakly interacting Bose gases~\cite{ho1999quasi,kimura2002breathing}, i.e., $\omega_\mathrm{b}^{1\mathrm{D}}=\sqrt{3}\omega_\mathrm{trap}$ and $\omega_\mathrm{b}^{2\mathrm{D}}=2\omega_\mathrm{trap}$.
Therefore, one may measure the accessible atomic sizes as well as the mode frequencies in cold atomic experiments, and probe the associated discontinuities to characterize this droplet-BEC phase transition of a dipolar quantum gas in this flattened external confinement.

\section{CONCLUSIONS AND OUTLOOKS} \label{sec:conclusions}
We have developed a comprehensive theoretical and numerical framework to investigate the structural properties and collective excitation dynamics of harmonically trapped dipolar quantum gases of lanthanide atoms such as $^{164}$Dy and $^{166}$Er.

By adopting a cylindrically symmetric Gaussian variational ansatz for the ground-state wave function and making a sum-rule analysis, we have successfully derived explicit analytical expressions for the axial and radial breathing-mode frequencies (i.e., $\omega_{\mathrm{b},z}, \omega_{\mathrm{b},\rho}$). These formulas yield upper bounds for low-lying breathing-mode frequencies without requiring explicit solutions of the full excitation spectrum.
Although our variational sum-rule approach captures overall trends, its quantitative accuracy depends on trap geometry owing to our decoupled variational ansatz for the wave function.
This sum-rule analysis is used to map out phase diagrams in the parameter space of the $s$-wave scattering length $a_s$, atom number $N$, and the trap aspect ratio $\lambda = \omega_z/\omega_\rho$. Two distinct regimes are probed, i.e.,
a dilute BEC phase with mode frequencies approaching the conventional values $\sqrt{3}\omega_{\mathrm{trap}}$ (one dimensional) or $2\omega_{\mathrm{trap}}$ (two dimensional), and a dense droplet phase characterized by significantly enhanced breathing frequencies. 
These analytical predictions of breathing-mode frequencies have been quantitatively validated by numerical solutions of the extended Gross-Pitaevskii equation, and the existing experimental measurements on $^{166}$Er~\cite{chomaz2016quantum} and $^{162}$Dy~\cite{tanzi2019supersolid}. The agreement is excellent in the droplet regime, while the sum-rule results provide an estimated upper bound in the BEC regime, as expected from the variational nature of the approach.
The transition between two phases over the range of scattering length can be either discontinuous first-order phase transitions or continuous smooth crossovers, depending on the trap aspect ratio. In particular, flattened or pancake-shaped traps favor first-order phase transitions with sharp discontinuities in peak density, cloud sizes, and breathing frequencies, whereas elongated or cigar-shaped traps yield smooth crossovers.

The analytical and numerical methodology presented in this work opens up several compelling pathways for future research in the field of dipolar quantum gases. 
While the Gaussian approximation in this work captures the essential physics, more sophisticated variational forms, e.g., with asymmetric widths or beyond-Gaussian correlations, could refine the predictions, especially near the critical point where the two minima merge. 

Besides, the enhanced breathing frequencies in the droplet regime and the clear discontinuity provide a clean experimental observable for probing the role of quantum fluctuations and the anisotropy of external confinement. The smooth crossover under an elongated trap has been clearly probed in the experiment, see Fig.~\ref{fig4:comparison-exps} (a), while the elusive phase transition under a typical flattened trap could be experimentally studied in the future where the discontinuous jump in breathing-mode frequency would be a hallmark signature.
In addition, instead of the linear response regime studied here, one may further explore nonlinear dynamics such as droplet collisions~\cite{ferrier2016liquid,ferioli2019collisions,hu2022collisional}, fragmentation, etc. Moreover, this paper focuses on the lanthanide atoms, and future work could extend to the recent polar molecules~\cite{bigagli2024observation,zhang2026observation} to test the universality of the theoretical framework.

\begin{acknowledgments}
This work is supported by the Science Challenge Project (Grant No. TZ2025017), the Natural Science Foundation of China (Grants No. 12474492, No. 12461160324, No. 12204413 and No. 12247101), the Science Foundation of Zhejiang Sci-Tech University (Grant No. 21062339-Y), the Fundamental Research Funds for the Central Universities (Grant No. lzujbky-2025-jdzx07), the Natural Science Foundation of Gansu Province (No. 25JRRA799), and the “111 Center” under Grant No. B20063. We acknowledge the use of high-performance computing clusters “Liguang II” at ZSTU.
\end{acknowledgments}

\emph{Data availability}---The data that support the findings of this article are openly available~\cite{zhang2026data}.

\appendix
\section{Explicit forms of the derivatives in the variational approach} \label{app:derivatives}
In this section, we present the explicit expressions for the derivatives in terms of two variables $\sigma_\rho$ and $\sigma_z$ used in the variational sum-rule analysis to derive the extreme points, as well as the breathing-mode frequencies in the main text. The explicit expressions are given as
\begin{widetext}
\begin{subequations}
\begin{eqnarray}
    \frac{\partial{f(\frac{\sigma_{\rho}}{\sigma_{z}})}}{\partial{\sigma_{\rho}}}&=&\frac{3\sigma_{\rho}}{(1-\frac{\sigma_{\rho}^2}{\sigma_{z}^2})^2\sigma_{z}^2}
    +\frac{4\sigma_{\rho}}{(1-\frac{\sigma_{\rho}^2}{\sigma_{z}^2})\sigma_{z}^2}
    +\frac{2\sigma_{\rho}(1+\frac{2\sigma_{\rho}^2}{\sigma_{z}^2})}{(1-\frac{\sigma_{\rho}^2}{\sigma_{z}^2})^2\sigma_{z}^2}
    -\frac{9\sigma_{\rho}^3 \mathrm{arctanh}\sqrt{1-\frac{\sigma_{\rho}^2}{\sigma_{z}^2}}}{(1-\frac{\sigma_{\rho}^2}{\sigma_{z}^2})^{\frac{5}{2}}\sigma_{z}^4}
    -\frac{6\sigma_{\rho} \mathrm{arctanh}\sqrt{1-\frac{\sigma_{\rho}^2}{\sigma_{z}^2}}}{(1-\frac{\sigma_{\rho}^2}{\sigma_{z}^2})^{\frac{3}{2}}\sigma_{z}^2},
\end{eqnarray}
\begin{eqnarray}
    \frac{\partial^2{f(\frac{\sigma_{\rho}}{\sigma_{z}})}}{\partial{\sigma_{\rho}^2}} &=&\frac{21\sigma_{\rho}^2}{(1-\frac{\sigma_{\rho}^2}{\sigma_{z}^2})^3\sigma_{z}^4}+\frac{16\sigma_{\rho}^2}{(1-\frac{\sigma_{\rho}^2}{\sigma_{z}^2})^2\sigma_{z}^4}+\frac{4}{(1-\frac{\sigma_{\rho}^2}{\sigma_{z}^2})\sigma_{z}^2}
    +\frac{8\sigma_{\rho}^2(1+\frac{2\sigma_{\rho}^2}{\sigma_{z}^2})}{(1-\frac{\sigma_{\rho}^2}{\sigma_{z}^2})^3\sigma_{z}^4}
    +\frac{9}{(1-\frac{\sigma_{\rho}^2}{\sigma_{z}^2})^2\sigma_{z}^2}
    +\frac{2(1+\frac{2\sigma_{\rho}^2}{\sigma_{z}^2})}{(1-\frac{\sigma_{\rho}^2}{\sigma_{z}^2})^2\sigma_{z}^2}
    \nonumber\\
    && -\frac{45\sigma_{\rho}^4 \mathrm{arctanh}\sqrt{1-\frac{\sigma_{\rho}^2}{\sigma_{z}^2}}}{(1-\frac{\sigma_{\rho}^2}{\sigma_{z}^2})^{\frac{7}{2}}\sigma_{z}^6}
    -\frac{45\sigma_{\rho}^2 \mathrm{arctanh}\sqrt{1-\frac{\sigma_{\rho}^2}{\sigma_{z}^2}}}{(1-\frac{\sigma_{\rho}^2}{\sigma_{z}^2})^{\frac{5}{2}}\sigma_{z}^4}
    -\frac{6 \mathrm{arctanh}\sqrt{1-\frac{\sigma_{\rho}^2}{\sigma_{z}^2}}}{(1-\frac{\sigma_{\rho}^2}{\sigma_{z}^2})^{\frac{3}{2}}\sigma_{z}^2},
\end{eqnarray}
\begin{eqnarray}
\frac{\partial{f(\frac{\sigma_{\rho}}{\sigma_{z}})}}{\partial{\sigma_{z}}} &=& -\frac{3\sigma_{\rho}^2}{(1-\frac{\sigma_{\rho}^2}{\sigma_{z}^2})^2\sigma_{z}^3}
    -\frac{4\sigma_{\rho}^2}{(1-\frac{\sigma_{\rho}^2}{\sigma_{z}^2})\sigma_{z}^3}
    -\frac{2\sigma_{\rho}^2(1+\frac{2\sigma_{\rho}^2}{\sigma_{z}^2})}{(1-\frac{\sigma_{\rho}^2}{\sigma_{z}^2})^2\sigma_{z}^3}
    +\frac{9\sigma_{\rho}^4 \mathrm{arctanh}\sqrt{1-\frac{\sigma_{\rho}^2}{\sigma_{z}^2}}}{(1-\frac{\sigma_{\rho}^2}{\sigma_{z}^2})^{\frac{5}{2}}\sigma_{z}^5}
    +\frac{6\sigma_{\rho}^2\mathrm{arctanh}\sqrt{1-\frac{\sigma_{\rho}^2}{\sigma_{z}^2}}}{(1-\frac{\sigma_{\rho}^2}{\sigma_{z}^2})^{\frac{3}{2}}\sigma_{z}^3},
\end{eqnarray}
\begin{eqnarray}
    \frac{\partial^2{f(\frac{\sigma_{\rho}}{\sigma_{z}})}}{\partial{\sigma_{z}^2}} &=&\frac{21\sigma_{\rho}^4}{(1-\frac{\sigma_{\rho}^2}{\sigma_{z}^2})^3\sigma_{z}^6}+\frac{16\sigma_{\rho}^4}{(1-\frac{\sigma_{\rho}^2}{\sigma_{z}^2})^2\sigma_{z}^6}+\frac{12\sigma_{\rho}^2}{(1-\frac{\sigma_{\rho}^2}{\sigma_{z}^2})\sigma_{z}^4}+\frac{8\sigma_{\rho}^4(1+\frac{2\sigma_{\rho}^2}{\sigma_{z}^2})}{(1-\frac{\sigma_{\rho}^2}{\sigma_{z}^2})^3\sigma_{z}^6}+\frac{15\sigma_{\rho}^2}{(1-\frac{\sigma_{\rho}^2}{\sigma_{z}^2})^2\sigma_{z}^4}+\frac{6\sigma_{\rho}^2(1+\frac{2\sigma_{\rho}^2}{\sigma_{z}^2})}{(1-\frac{\sigma_{\rho}^2}{\sigma_{z}^2})^2\sigma_{z}^4}\nonumber\\
    && -\frac{45\sigma_{\rho}^6 \mathrm{arctanh}\sqrt{1-\frac{\sigma_{\rho}^2}{\sigma_{z}^2}}}{(1-\frac{\sigma_{\rho}^2}{\sigma_{z}^2})^{\frac{7}{2}}\sigma_{z}^8}
    -\frac{63\sigma_{\rho}^4 \mathrm{arctanh}\sqrt{1-\frac{\sigma_{\rho}^2}{\sigma_{z}^2}}}{(1-\frac{\sigma_{\rho}^2}{\sigma_{z}^2})^{\frac{5}{2}}\sigma_{z}^6}
    -\frac{18\sigma_{\rho}^2 \mathrm{arctanh}\sqrt{1-\frac{\sigma_{\rho}^2}{\sigma_{z}^2}}}{(1-\frac{\sigma_{\rho}^2}{\sigma_{z}^2})^{\frac{3}{2}}\sigma_{z}^4}.
\end{eqnarray}
\end{subequations}
\end{widetext}
In Eqs.~\eqref{eq:omegaB-z} and ~\eqref{eq:omegaB-rho}, the subscript “0” in the derivatives of the function $f(x)$ denotes the evaluation of the above expressions by substituting the extremum $(\sigma_{\rho_0},\sigma_{z_0})$.

\section{Comparisons of collective mode frequencies between the time-dependent extended GPE and the Bogoliubov–de Gennes equations} \label{app:GPEvsBdG}

In this section, we present the mode frequencies obtained by introducing specific weak perturbations in the numerical simulations of the time-dependent extended GPE in Sec.~\ref{sec:extendedGPE}. These results are further compared with the existing full spectra calculated from the Bogoliubov–de Gennes (BdG) equations in Refs.~\cite{baillie2017collective,blakie2023axial}, and the analytical mode frequency in Eq.~\eqref{eq:omegaB-z} estimated using the sum-rule approach with a Gaussian variational ansatz.
\begin{figure}[t]
\centering
\includegraphics[width=0.48\textwidth]{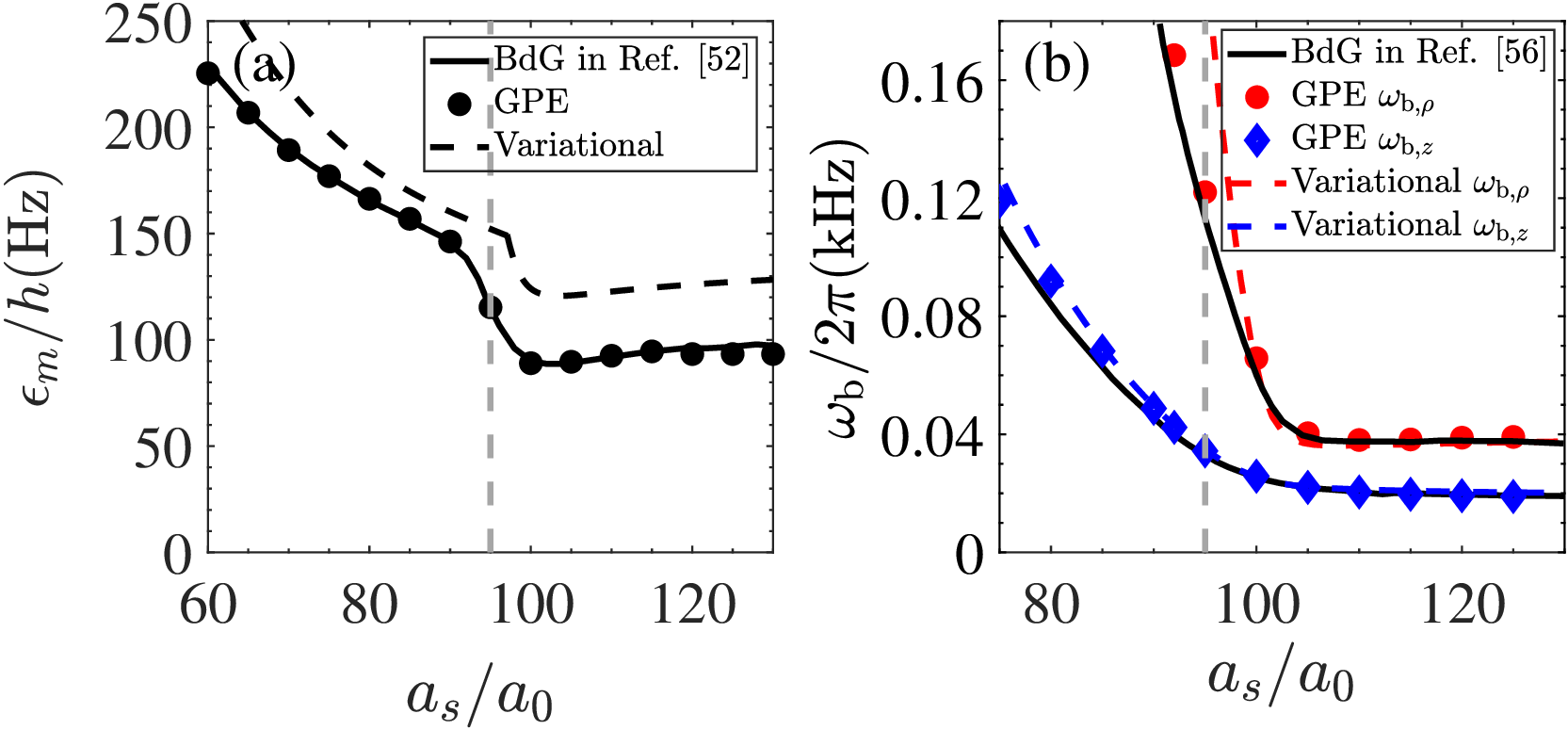}
\caption{\justifying
Comparison of breathing-mode frequencies calculated by the numerical time-dependent extended GPE together with specific slight perturbations (i.e., solid dots and diamonds) and the sum-rule approach within the variational approximation (i.e., dashed lines) adopted in this work, with the existing results obtained from the Bogoliubov–de Gennes equations (i.e., solid lines).
The breathing-mode energy $\epsilon_m$ or frequencies $\omega_{\mathrm{b}}$ are shown as a function of the $s$-wave scattering length $a_s$ for (a) $2\times10^4$ $^{164}$Dy atoms in a spherical trap with $\omega=2\pi\times70$ Hz as studied in Ref.~\cite{baillie2017collective}, and (b) $10^4$ $^{164}$Dy atoms in traps with $(\omega_x=\omega_y,\omega_z)=2\pi\times(20,10)$ Hz as studied in Ref.~\cite{blakie2023axial}, respectively. The vertical gray dashed lines indicate the estimated critical points separating the droplet and BEC regimes in the two references, both of which occur at approximately $a_s\approx95a_0$. }
\label{fig7:GPEvsBdG}
\end{figure}

In Fig.~\ref{fig7:GPEvsBdG}(a), the breathing-mode frequency of approximately $2\times 10^4$ $^{164}$Dy atoms in a spherical trap with $\omega=2\pi \times 70$ Hz is presented as a function of the scattering length $a_s$, spanning from the left droplet regime to the right BEC regime.
The solid black line denotes the lowest $m=0$ breathing-mode frequency, extracted from the full BdG spectrum as done in Ref.~\cite{baillie2017collective}.
As expected, the mode frequency calculated by the time-dependent extended GPE in this work, shown as solid dots, exhibits excellent agreement with the BdG results. In contrast, the breathing-mode frequency (i.e., black dashed line) predicted by the sum-rule approach based on a Gaussian variational ansatz exhibits the same qualitative behavior, but with a systematic overestimation.
Similarly, in Fig.~\ref{fig7:GPEvsBdG}(b), the radial and axial breathing-mode frequencies of $10^4$ $^{164}$Dy atoms in a cylindrically symmetric trap with $(\omega_x=\omega_y,\omega_z)=2\pi\times(20,10)$ Hz are displayed as functions of $a_s$. The radial and axial breathing-mode frequencies obtained from the extended GPE, namely $\omega_{\mathrm{b},\rho}$ (i.e., red dots) and $\omega_{\mathrm{b},z}$ (i.e., blue diamonds), again show quantitative agreement with the BdG predictions (i.e., solid black lines) reported in Ref.~\cite{blakie2023axial}. The low-energy breathing-mode frequencies (i.e., red and blue dashed lines) predicted by the sum-rule approach coincide with the numerical results in the BEC regime (to the right of the vertical gray dashed line at $a_s\approx95a_0$), while they are slightly overestimated in the droplet regime on the left.

\bibliography{breathingDipolar}

\end{document}